\documentclass[11pt,a4paper]{article}
\pdfoutput=1

\usepackage{bm}
\usepackage{amsmath,amssymb}
\usepackage{cite}

\usepackage{graphicx}
\usepackage{color}

\usepackage{array} %,booktabs}
\usepackage{hyperref} %%put this package declaration last to make work since it redefines things!!%%
\hypersetup{colorlinks=true, citecolor=blue, filecolor=black, linktocpage=true,  linkcolor=blue, urlcolor=blue, pdfpagemode=UseNone}
\numberwithin{figure}{section}
\numberwithin{equation}{section}

\newcommand{\be}{\begin{equation}}
\newcommand{\ee}{\end{equation}}
\newcommand{\bea}{\begin{eqnarray}}
\newcommand{\eea}{\end{eqnarray}}

\usepackage[normalem]{ulem}

\newcommand\ie{\textit{i.e.}\ }
\newcommand\eg{\textit{e.g.}\ }
\newcommand\cf{\textit{cf.}\ }

\newcommand{\viz}{{\it viz.}\ }

\newcommand{\bi}{\begin{itemize}}
\newcommand{\ei}{\end{itemize}}
\def\beal#1\eeal{\begin{align}#1\end{align}}

\newcommand{\vp}{\varphi}
\newcommand{\nfrac}[2]{\,#1/#2}

\def\b{\beta}

\def\e{\epsilon}

\def\e{\epsilon}

\usepackage{chngcntr}
\newcounter{para}

\begin{document}

\begin{titlepage}
%\begin{flushright}
%%{\tt hep-ph/yymmnn}
%{\tt SHEP xx-xx}
%\end{flushright}

\begin{center}
{\huge \bf Asymptotic solutions in asymptotic safety} 
%\vskip.3cm
%{\huge \bf  and etc} 
\end{center}
\vskip1cm

%\title{xxx}
%\author{Tim R. Morris}

\begin{center}
{\bf Sergio Gonzalez-Martin,$^a$ Tim R. Morris$^b$ and Zo\"e H. Slade$^b$}
\end{center}

%\affiliation{
\begin{center}
{$^a$ \it Departamento de F\'isica Te\'orica and Instituto de F\'{\i}sica Te\'orica (IFT-UAM/CSIC),\\
	Universidad Aut\'onoma de Madrid, Cantoblanco, 28049, Madrid, Spain.}\\
{$^b$ \it STAG Research Centre \& Department of Physics and Astronomy,\\  University of Southampton,
Highfield, Southampton, SO17 1BJ, U.K.}\\
\vspace*{0.3cm}
{\tt sergio.gonzalez.martin@csic.es, T.R.Morris@soton.ac.uk,\\ Z.Slade@soton.ac.uk}
\end{center}

\abstract{We explain how to find the asymptotic form of fixed point solutions in functional truncations, in particular $f(R)$ approximations. We find that quantum fluctuations do not decouple at large $R$, typically leading to elaborate asymptotic solutions containing several free parameters. By a counting argument, these can be used to map out the dimension of the fixed point solution spaces. They are also necessary to validate the numerical solution, and provide the physical part in the limit that the cutoff is removed:  the fixed point equation of state. As an example we apply the techniques to a recent $f(R)$ approximation by Demmel \textit{et al}, finding asymptotic matches to their numerical solution. Depending on the value of the endomorphism parameter, we find
many other asymptotic solutions and fixed point solution spaces of differing dimensions,  yielding several alternative scenarios for the equation of state. Asymptotic studies of other $f(R)$ approximations are needed to clarify the picture.}

\end{titlepage}

\tableofcontents

%\newpage

\section{Introduction}
\label{sec:Intro}

The asymptotic safety approach to finding a quantum theory of gravity relies on finding a non-Gaussian ultraviolet fixed point of the gravitational renormalization group flow \cite{Weinberg:1980,Reuter:1996}. This flow is couched in terms of a Wilsonian effective action \cite{Wilson:1973}, using the reformulation of the exact renormalization group  in terms of an effective average action $\Gamma_k$, the Legendre effective action with an infrared cutoff scale $k$ \cite{Nicoll1977,Wetterich:1992,Morris:1993}. The existence of such a fixed point has been investigated and supported in a substantial number of approximations. For reviews and introductions see \cite{Reuter:2012,Percacci:2011fr,Niedermaier:2006wt,Nagy:2012ef,Litim:2011cp} and for some of the many recent approaches see refs. \cite{Demmel:2013myx,Demmel2015b,Ohta:2015efa,Ohta2016,Percacci:2016arh,Morris:2016spn,Falls:2016msz,Ohta:2017dsq,Percacci:2015wwa,Labus:2015ska,Eichhorn:2015bna,Dietz:2016gzg,Falls:2016wsa, Denz:2016qks,Christiansen:2016sjn,Eichhorn:2016vvy,Biemans:2016rvp,Gies:2016con,Morris:2016nda,Wetterich:2016ewc,Henz:2016aoh,Labus:2016lkh,Falls:2017cze,Hamada:2017rvn,Christiansen:2017gtg,Biemans:2017zca}.

In order to study this, it is necessary to work within some non-perturbative approximation scheme. %it is still the case that t
The vast majority of these approximations are still formulated by keeping only a finite number of couplings in the effective action. A weakness of such an approach is that fixed points are effectively the solutions of polynomial equations in these couplings, which thus only allow for discrete solutions. But physical systems exist with lines or even higher dimensional surfaces of fixed points, parametrised by exactly marginal couplings (in supersymmetric theories these are common and termed moduli). Furthermore lines and planes of fixed points have also been found in approximations to asymptotic safety\footnote{and in a perhaps related approximation in scalar-tensor gravity \cite{Benedetti:2013nya}}  
\cite{Dietz:2012ic,Dietz:2013sba,Dietz:2016gzg}.
For isolated fixed points, careful treatment of  polynomial approximations taken to high order, can allow extraction of convergent results, however one does not see in this way the singularities at finite field or asymptotic behaviour at diverging field, 
which are actually responsible for determining their high order behaviour, through the analytic structure they impose \cite{Morris:1994ki}. In fact such large field effects can invalidate deductions from polynomial truncations \cite{Morris:1994ki,Morris:1995he,Morris:1996nx} and/or restrict or even exclude the existence of global solutions
%determine the analytic structure of the solutions and thus the large order behaviour of the fixed point solutions to polynomial approximations \cite{Morris:1994ki}, but also restrict or even invalidate them \cite{Morris:1996nx}. 
\cite{Morris:1994ie,Morris:1994jc,Morris:1997xj,Morris:1998}. A good example is provided by some of the most impressive evidence for asymptotic safety to date, polynomial expansions in scalar curvature $R$ including all powers up to $R^{34}$ \cite{Falls:2013bv,Falls:2014tra,Falls:2016wsa},
which are however derived from a differential equation for an $f(R)$ fixed point Lagrangian
\cite{Codello:2008} which was shown in ref. \cite{Dietz:2012ic} to have no global solutions, as a consequence of fixed singularities at finite field.

Therefore if we are to believe that a putative isolated fixed point is not just an artifact of an insufficient approximation, we must go beyond polynomial truncations to approximations that keep an infinite number of couplings. Arguably the simplest such approximation is to keep a full function $f(R)$, making the ansatz:
\be
\label{ansatz EAA}
\Gamma_k[g]=\int\,d^4x\sqrt{g}\, f(R)\,,
\ee
and to date this is the only such approximation that has been investigated \cite{Machado:2007,Codello:2008,Benedetti:2012,Demmel:2012ub,Demmel:2013myx,Benedetti:2013jk,Demmel:2014hla,Demmel2015b,Ohta:2015efa,Ohta2016,Percacci:2016arh,Morris:2016spn,Falls:2016msz,Ohta:2017dsq}, together with some closely related approximations in scalar-tensor 
\cite{Percacci:2015wwa,Labus:2015ska} and unimodular \cite{Eichhorn:2015bna} gravity, and in three space-time dimensions \cite{Demmel:2014fk}.
It should be emphasised that this actually goes beyond  keeping a countably infinite number of couplings, the Taylor expansion coefficients $g_n=f^{(n)}(0)$, because \textit{a priori} the large field parts of $f(R)$ contain degrees of freedom which are unrelated to all these $g_n$.

The result of such an approximation is a fixed point equation which is either (depending on the implementation of the effective cutoff $k$) a third order or a second order, non-linear ODE (ordinary differential equation)  for the dimensionless function $\vp(r)$, where 
\be
\label{dim vars}
r \equiv {R}\, k^{-2}\,, \qquad\quad f({R}) \equiv k^4 \vp({R}k^{-2})\,.
\ee
To be concrete we will give the discussion assuming the typical case where the equation is derived on a space of positive curvature (effectively the Euclidean four-sphere, the discussion is readily adapted to negative curvatures) in which case a fixed point corresponds to a smooth global solution $\vp(r)$ over the domain $r\in[0,\infty)$. Now, to understand the solutions of these equations both physically and mathematically, it is crucial to develop the asymptotic solutions $\vp_{asy}(r)$, as we explain below. 

Although these ODEs are  complicated, \eg \eqref{fp}, the asymptotic solutions $\vp_{asy}(r)$ can fortunately be  found analytically and in full generality \cite{Dietz:2012ic,Dietz:2016gzg} by adopting techniques developed much earlier for scalar field theories \cite{Morris:1994ki,Morris:1994ie,Morris:1994jc}. These techniques apply to any functional truncation of the exact renormalization group fixed point equations, such that the result is an ODE or coupled set of ODEs (as \eg in \cite{Dietz:2016gzg}), although to be concrete we will focus on solving for 
$\vp_{asy}(r)$.
Perhaps because these techniques were covered only briefly and without outlining the general treatment, they have yet to be entirely adopted, meaning that the functional solution spaces for many of the formulations \cite{Percacci:2015wwa,Labus:2015ska,Eichhorn:2015bna,Demmel:2014fk,Demmel:2012ub,Demmel:2013myx,Benedetti:2013jk,Demmel:2014hla,Demmel2015b,Ohta:2015efa,Ohta2016,Percacci:2016arh,Falls:2016msz,Ohta:2017dsq} remain unexplored or at best only partially explored. The main purpose of the present paper is to improve this situation, by  describing in detail and with as much clarity as possible how the techniques allow to  fully unfurl the asymptotic solutions. As an illustration we choose to apply these techniques to one example of a fixed point ODE \cite{Demmel2015b} which fortuitously provides a zoo of asymptotic solutions of different types. 

%In sec. \ref{sec:conclusions} we broaden out the discussion to include some other formulations and try to draw some more general conclusions.

Perhaps another reason why the asymptotically large $r=R/k^2$ region may have been under-explored is that it has not been clear what meaning should be attached to this region when $1/k$ is larger than the physical size $1/\sqrt{R}$ of the manifold, despite the fact that we know that the infrared cutoff $k$ is artificial and introduced by hand and the physical effective action,
\be 
\label{phys-eff-ac}
\Gamma=\lim_{k\to0} \Gamma_k\,,
\ee
is therefore only recovered when the cutoff is removed. This puzzle was brought into sharp relief in formulations that have a gap, \ie a lowest eigenvalue which is positive, so that large $r$ then corresponds to $k$ being less than any eigenvalue  \cite{Demmel:2014fk,Demmel2015b,Ohta2016,Falls:2016msz}. This issue was recently resolved in ref. \cite{Morris:2016spn} where it was shown to be intimately related to ensuring background independence (but in a way that can be resolved even for single-metric approximations, which is just as well since only the refs. \cite{Morris:2016spn,Percacci:2016arh, Ohta:2017dsq} in the list above  actually go beyond this approximation). Wilsonian renormalization group concepts do not apply to a single sphere. In particular, 
%in order for the renormalization group concepts to make sense, we now know that solutions should be globally defined in this sense, whether or not a smoothing procedure is applied \cite{Morris:2016spn}. We recall that, 
although in these formulations, $k$ can be low enough on a sphere of given curvature $R$ that there are no modes left to integrate out, the fixed point equation should be viewed as summarising the state of a continuous ensemble of spheres of different curvatures. From the point of view of the ensemble there is nothing special about the lowest mode on a particular sphere. The renormalization group should be smoothly applied to the whole ensemble, and it is for this reason that one must require that smooth solutions exist over the whole domain $0\le r<\infty$.

In the sec. \ref{sec:overview}, we introduce the ODE we will study, provide a compendium of our results, and discuss their meaning.
 In the secs. \ref{sec:pow-laws} and \ref{sec:non-pow} we provide the details of how these are derived. We finish this introduction,  by listing reasons why the asymptotic solution is so important.
 
 \subsection{Quantum fluctuations do not decouple}
\label{sec:quantum}

%These last two are physical reasons why $\vp_{asy}(r)$ is so important. 
In the application to scalar field theory  \cite{Morris:1994ki,Morris:1994ie,Morris:1994jc,Morris:1998}, the leading asymptotic behaviour was always found by neglecting the right-hand side of the fixed point equation (more generally flow equation). This made physical sense since the right-hand side encodes the quantum fluctuations, and at large field one would expect that these are negligible in comparison. Therefore the asymptotic solution simply encodes the passage to mean field scaling, characteristic of the classical limit. We find that with functional approximations to quantum gravity, the situation is radically different. The leading asymptotic solution $\vp_{asy}(r)$ intimately depends on the right-hand side and never on the left-hand side alone. We will see this for the large $r$ solutions of the fixed point equation, \eqref{fp}, derived by Demmel \textit{et al} \cite{Demmel2015b}. This behaviour confirms what we already found in a different $f(R)$ approximation in ref. \cite{Dietz:2012ic}, and also for a conformal truncation in ref. \cite{Dietz:2016gzg}. 
%and the solutions for \eqref{fp} \cite{Demmel2015b} in sec. \ref{sec:overview}).  
Again this actually makes physical sense because the analogue here of large field is large curvature which therefore shrinks the size of the space-time and thus forbids the decoupling of quantum fluctuations. In fact by Heisenberg's uncertainty principle we must expect  that the quantum fluctuations become ever wilder. We note  that it is the conformal scalar contribution that is determining the leading behaviour \cite{Dietz:2012ic,Demmel2015b,Dietz:2016gzg} and appears to be related to the so-called conformal instability\cite{Gibbons:1978ac,Dietz:2012ic,Dietz:2016gzg}. In any case,
we see that for quantum gravity, the asymptotic solution $\vp_{asy}(r)$ encodes the deep non-perturbative quantum regime.

\subsection{Physical part}
\label{sec:intro-phys}

The asymptotic solution contains the only physical part of the fixed point effective action. Recall that the effective infrared cutoff $k$ is added by hand and the physical Legendre effective action \eqref{phys-eff-ac} is recovered only in the limit that this cutoff is removed. This is of course done while holding the physical quantities (rather than say scaled quantities) fixed. In normal field theory, \eg scalar field theory, the analogous object is the universal scaling equation of state, which for a constant field precisely at the fixed point takes the simple form 
\be 
\label{Vscale}
V(\vp) = A \,\vp^{d/d_\vp}\,,
\ee 
where $d$ is the space-time dimension and $d_{\vp}$ is the full scaling dimension of the field (\ie incorporating also the anomalous dimension). In the current case we keep fixed the the constant background scalar curvature $R$. Thus by \eqref{ansatz EAA}, the only physical part of the fixed point action in this approximation is:
\be 
\label{phys}
f(R) |_{\text{phys}} = \lim_{k\to 0} k^4\, \vp(R/k^2) = \lim_{k\to 0} k^4\, \vp_{asy}(R/k^2)\,.
\ee
The significance of this object is further discussed in sec. \ref{sec:phys}, in the light of the results we uncover.

\subsection{Dimensionality of the fixed point solution space}
\label{sec:dims}

For given values of the parameters, the fixed point ODEs are too complicated to solve analytically,\footnote{although special analytical solutions were found by tuning the endormorphism parameters \cite{Ohta2016}.} and challenging to solve numerically. 
However the dimension, $d_{FP}$, of their solution space, namely whether the fixed points are discrete, form lines, or planar regions \textit{etc.}, can be found by inspecting the fixed singularities and the asymptotic solutions. 

To see this we express the fixed point ODE in normal form by solving for the highest derivative:
\be
\label{norm}
\vp^{(n)}(r) = rhs\, ,
\ee
where $n=n_{ODE}$ is the order of the ODE, and $rhs$ (right-hand side) contains only rational functions of $r$ and lower order differentials\footnote{including $\vp$ itself \ie $m=0$} $\vp^{(m<n)}(r)$.
The fixed singularities are found at points $r=r_i$ where this expression develops a pole for generic $\vp^{(m<n)}(r)$. For the solution to pass through the pole requires a boundary condition relating the $\vp^{(m<n)}(r)$, one for each pole. By a fixed singularity, we will mean one of these poles.

At the same time such non-linear ODEs suffer moveable singularities, points where $rhs$ diverges as a consequence of specific values for the  $\vp^{(m<n)}(r)$. The number of these that operate in practice depends on the solution itself. However if the solution is to exist globally then it exists also for large $r$, where we can determine it analytically in the form of its asymptotic solution $\vp_{asy}(r)$, this being the central topic of this paper. The 
number of constraints implicit in $\vp_{asy}(r)$ is equal to $n_{ODE}-n_{asy}$,
where $n_{asy}$ is the number of free parameters in $\vp_{asy}(r)$. This can be seen straightforwardly by noting that the maximum possible number of free parameters is $n_{asy}=n_{ODE}$; if $\vp_{asy}(r)$ contains any less then this implies that there are $n_{ODE}-n_{asy}$ relations between the $\vp^{(0\le m\le n)}_{asy}$ at any large enough $r$, which may be used as boundary conditions. Now the number of moveable singularities that operate for a solution with these asymptotics is also equal to $n_{ODE}-n_{asy}$, providing we have uncovered the full set of free parameters in the asymptotic solution, as has been explicitly verified by now in many cases \cite{Morris:1994ki,Morris:1994ie,Morris:1994jc,Morris:1995he,Morris:1996nx,Morris:1997xj,Morris:1998,Dietz:2012ic,Bridle:2013sra,Dietz:2016gzg}. This follows because the moveable singularities can also occur at large $r$ where they influence the form of $\vp_{asy}$. Indeed linearising the ODE about $\vp_{asy}$, the perturbations can also be solved for analytically. The missing free parameters in $\vp_{asy}$ correspond to perturbations that grow faster than $\vp_{asy}$, overwhelming it and invalidating the assumptions used to derive it in the first place. These perturbations can be understood to be the linearised expressions of these moveable singularities \cite{Morris:1994ki,Morris:1994ie,Morris:1994jc}. To summarise, if the number of fixed singularities operating in the solution domain is $n_{s}$, then the dimension of the solution space is simply given by
\be 
\label{counting}
d_{FP} = n_{ODE} - n_{s} - (n_{ODE}-n_{asy}) = n_{asy} - n_{s}\,,
\ee
where $d_{FP}=0$ indicates a discrete solution set which may or may not be empty, 
and $d_{FP}<0$ corresponds to being overconstrained, \ie having no solutions.
In sec. \ref{sec:dimensionality-ef}, we will illustrate this by working out the dimension of the fixed point solutions for eleven of the possible asymptotic behaviours. We discuss their significance in sec. \ref{sec:which}.

At the same time the counting argument \eqref{counting} aids in the numerical solution. For example it tells us where it is hopeless to look for global numerical solutions, namely where $d_{FP}<0$, and to improve the numerical accuracy if the numerical solution apparently enjoys more free parameters than allowed by $d_{FP}$  \cite{Dietz:2012ic}. 

\subsection{Validation of the numerical solution}
\label{sec:validate}

\textit{A priori} one might think that the analytical solutions for $\vp_{asy}(r)$ can be dispensed with in favour of a thorough numerical investigation. The problem is that without knowledge of $\vp_{asy}(r)$, there is no way to tell whether the numerical solution that is found is a global one, \viz exists and is smooth as $r\to\infty$, or will ultimately end at some large $r$ in a moveable singularity. In fact if the numerical solution is accessing a regime where the number of free asymptotic parameters $n_{asy}<n_{ODE}$, it will actually prove impossible to integrate numerically out to arbitrarily large $r$. Instead the numerical integrator is guaranteed to fail at some critical value. The reason is that it requires infinite accuracy to avoid including one of the linearised perturbations that grow faster than $\vp_{asy}$ which as we said, signal that the solution is about to end in a moveable singularity. On the other hand, if one can extend the solution far enough to provide a convincing fit to the analytical form of $\vp_{asy}(r)$, then one confirms with the requisite numerical accuracy that the numerical solution has safely reached the asymptotic regime
\cite{Dietz:2012ic,Bridle:2013sra,Dietz:2016gzg}, after which its existence is established, and its form is known, over the whole domain. We will see an example of this in sec. \ref{sec:numer-pow} where we will see in fact that the numerical solution found in ref. \cite{Demmel2015b} matches the power-law asymptotic expansion \eqref{full-pow-beta1/6}, but such that it would need to be integrated out twice as far in order to be sure of its asymptotic fate.

\section{Overview}
\label{sec:overview}

\subsection{Fixed point equation}

The fixed point equation that we will be studying is given by \cite{Demmel2015b}:
\be
\label{fp}
4\vp-2r\vp'=\frac{\tilde{c}_1\vp'-2\tilde{c}_2 r\vp''}{3\vp-(3\alpha r + r - 3)\vp'}+\frac{c_1\vp' + c_2 \vp'' - 2c_4 r \vp'''}{(3\beta r + r -3)^2\vp''+(3-(3\beta+2)r)\vp' + 2\vp}\,,
\ee
where $\vp(r)$ is defined in \eqref{dim vars}, and prime indicates differentiation with respect to $r$. This gives the scaled dependence on the curvature of a Euclidean four-sphere. Therefore we are searching for smooth solutions defined on the domain $0\le r<\infty$. Each such solution is a fixed point of the renormalization group flow.

The coefficients $\tilde{c}_i$ and $c_i$ depend on $r$ and the endomorphism parameters\footnote{The endomorphisms are curvature terms added to the covariant Laplacian, allowing flexibility in how its eigenvalues are integrated out \cite{Demmel2015b}. They are further discussed in secs. \ref{sec:which} and \ref{sec:conclusions}.} $\alpha$ and $\beta$, and are given as
\begin{align}
\tilde{c}_1 &=-\frac{5 (6 \alpha  r+r-6) \left(\left(18 \alpha ^2+9 \alpha -2\right) r^2-18 (8 \alpha +1) r+126\right)}{6912 \pi ^2}\nonumber\,,\\
\tilde{c}_2 &=-\frac{5 (6 \alpha  r+r-6) ((3 \alpha +2) r-3) ((6 \alpha -1) r-6)}{6912 \pi ^2}\,,\label{coeffs-c}\\
c_1&=-\frac{((6 \beta -1) r-6) \left((6 \beta -1) \beta  r^2+(10-48 \beta ) r+42\right)}{2304 \pi^2}\,,\nonumber\\
c_2&=-\frac{((6 \beta -1) r-6) \left(\left(54 \beta ^2-3 \beta -1\right) \beta  r^3+\left(270 \beta ^2+42 \beta -35\right) r^2-39 (18 \beta +1) r+378\right)}{4608 \pi ^2}\,,\nonumber\\
c_4&=\frac{(\beta  r-1) ((6 \beta -1) r-6)^2 ((9 \beta +5) r-9)}{4608 \pi ^2}\,.\nonumber
\end{align}
In ref. \cite{Demmel2015b}, the authors set
%\be
%\label{ELE}
$\alpha=\beta-2/3$ %\, ,\ee
and we will do the same.\footnote{This sets the lowest eigenvalues equal for the scalar and tensor modes. Following ref. \cite{Morris:2016spn}, see also above sec. \ref{sec:dims},
it is not clear what significance should be attached to this however.}

\subsection{Fixed singularities}
\label{sec:fixed-sing-count}

The ODE \eqref{fp} is third order and thus admits a three-parameter set of solutions locally.
As reviewed in \ref{sec:dims}, the fixed singularities will limit this parameter space. The positions of the fixed singularities are determined by casting the flow equation into normal form \eqref{norm} (with $n=3$). The zeroes of the coefficient $c_4$ then give the points where the flow equation develops a pole. These poles are given by \cite{Demmel2015b}:
\be
r_1=0\,,\qquad\quad r_2=\frac{9}{5+9\beta}\,, \qquad\quad r_3=\frac{1}{\beta}\,, \qquad\quad r_{4,5}=\frac{6}{6\beta-1}\,.\nonumber
\ee
Note that there is a double root $r_{4,5}$ and that the root $r_1$, which is actually there for good physical reasons \cite{Benedetti:2012,Dietz:2012ic}, is always present, whereas the positions of the last 4 roots depend on the value $\beta$ takes. Different choices for $\beta$ will result in a different number of fixed singularities being present in the range $r\geq 0$, as shown in table \ref{table: fixed sings}.
\begin{table}
\begin{center}\begin{tabular}{ c|c }
Range of $\beta$ & Singularities\\
\hline
$1/6 < \beta$ & $r_1,\,r_2,\,r_3,\,r_{4,5}$ \\ 
$0<\beta \leq 1/6$ & $r_1, \,r_2,\,r_3$ \\ 
$-5/9<\beta\leq0$ & $r_1,\, r_2$ \\ 
$\beta\leq-5/9$ & $r_1$\\
\end{tabular}
\end{center}
\caption{List of fixed singularities present for different choices for $\beta$.}
\label{table: fixed sings}
\end{table}
If no additional constraints emerge from the asymptotic behaviour of the solution then choosing $0<\beta\leq 1/6$ leads by \eqref{counting} to $n_{FP} =3-n_s=0$, which means that isolated fixed point solutions (or no solutions) can be expected.
%3 constraints ensuring an isolated and globally defined solution to the flow equation. 
The authors of \cite{Demmel2015b} choose $\beta = 1/6$ for this reason. It is also noted in \cite{Demmel2015b} that in addition this choice simplifies the numerical analysis.  We will analyse the fixed point equation for general $\beta$, both to uncover the extent to which the results depend on the particular choice and to demonstrate and explain the asymptotic methods in a large variety of examples. But since $\beta=1/6$ was chosen in ref. \cite{Demmel2015b}, we will pay special attention to this value.

If $\beta>1/6$ is chosen then $n_{FP}<0$, the ODE is overconstrained and global solutions do not exist. On the other hand non-positive $\beta$ give rise to continuous sets, again assuming that no extra constraints are coming from the solution at infinity. For example $-5/9<\beta\leq0$, would give rise to only 2 fixed singularities, resulting in a one-parameter set of solutions \ie a line of fixed points, while $\beta\le-5/9$ gives us a plane of fixed points. 

\subsection{Asymptotic expansions}
\label{sec:asymptotics-overview}

We now list all the asymptotic solutions that we found. 
They can have up to three parameters, which are always called $A$, $B$ and $C$.

%\subsection{Power law asymptotics}

%We find the following asymptotic solutions whose leading term is a power of $r$.

\paragraph{(a)} As covered in sec. \ref{Leading behaviour}, there exists a power-law solution where the leading power is $r^0$. It takes the form
\be 
\label{n=0}
\vp(r)= A + k_1/r^2+\cdots\,,
\ee
for all $\beta\ne0$, where $k_1$ is given by \eqref{n=0 k1}. 

\paragraph{(b)} For $\beta=0$, the subleading power is altered and the solution changes to 
\be
\label{n=0no2}
\vp(r)=A -\frac{18432\,{\pi }^{2}{A}^{2}}{535\,r} +\cdots\,.
\ee
as explained in sec. \ref{sec:excep-numer}. At this value of $\beta$ only this asymptotic solution is allowed.

\paragraph{(c)} As covered in sec. \ref{Leading behaviour}, for $n$ a root of \eqref{double root soln} such that $n<2$, the asymptotic solution
\be 
\label{n general subleading}
\vp(r)= Ar^n + k_1r^{2n-2}+\cdots\,,
\ee
with $k_1$ given by \eqref{n general k1}, exists for all $\beta \notin (-0.4835,-0.4273)$, except as explained in sec. \ref{sec:excep-numer} for $\beta=1/6$ and $\beta=0$, and except for the values
\be
\label{beta n=2}
\beta = \beta_\pm := \frac{3}{13} \pm \frac{\sqrt{285}}{78} = 0.01433,\ 0.4472 \,,
\ee
as explained at the end of sec. \ref{Leading behaviour}.  
When $n$ is complex, which happens for $\beta\in(-1.326,-0.4474)$ the parameter $A$ is in general also complex and the real part of \eqref{n general subleading}
should be taken leading to $\vp(r) \sim r^{\text{Re}(n)} \sin(\text{Im}(n)\log r +B)$ type behaviour. The values $n$ are plotted in fig. \ref{scenario 2}.

\paragraph{(d)} As explained at the end of sec. \ref{Leading behaviour}, for the values \eqref{beta n=2}, $n=2$ is a root of \eqref{double root soln}, however the asymptotic solution is not given by \eqref{n general subleading} but
\be
\label{n=2sol}
\vp(r) = Ar^2+k_1r+\cdots,
\ee
where $k_1$ is given by \eqref{n=2sol k1}.

\paragraph{}The techniques set out in secs. \ref{Missing parameters} and \ref{missing params 2} need to be followed to find the missing parameters in the above solutions (a) -- (d), before we can discover the dimension $n_{FP}$ of their corresponding solution space. Of course we already know from table \ref{table: fixed sings} that there are no solutions (\ie $n_{FP}<0$) for $\beta>1/6$. In the remaining asymptotic solutions we also uncover the missing parameters.

\paragraph{(e)} For generic $\beta$ the following solution:
\be
\label{solution}
\vp(r) = \vp_{pow}(r):= A\,r^{3/2} + k_1 \, r + k_2 \, r^{1/2} + k_3\, \log\left(\frac{r}{b}\right)
+k_4\frac{\log\left(\frac{r}{b}\right)}{\sqrt{r}}+ \frac{k_5}{\sqrt{r}} +\cdots\,,
\ee
where $B=\log b$ is a second free asymptotic parameter,
forms the basis for the asymptotic solutions below. As  explained in sec.  \ref{Exceptions} it fails to exist for $\beta=0$, 1, and $\pm1/\sqrt{27}$. The leading part is derived in sec. \ref{Leading behaviour} and the subleading parts in sec. \ref{Sub-leading behaviour}. The subleading coefficients are functions of $A$ and $\beta$, where $k_1$ is given in \eqref{Fval} and the others are given in appendix \ref{AppendixA}. As explained in sec. \ref{exceptions}, exceptions develop at poles of these subleading coefficients where the corresponding term and subleading terms then develop an extra $\log r$ piece. As shown in sec. \ref{Missing parameters}, the full asymptotic solution  is then one of  the following forms:
\beal
&\vp_{pow}(r) \,,  &&\beta\in (-\infty,-0.1809)\cup(0.1931,0.4042)\cup(0.8913,\infty)\backslash\{-\frac59\}\,,\label{p3pos}\\
&\vp_{pow}(r) + C\, r^{p_3 +\frac{3}{2}}+\cdots\,,  &&\beta\in(-0.1809,0.1931)\cup(0.4042,0.8913)\backslash\{\frac16\}\,,\label{p3neg}\\
&\vp_{pow}(r) + C\, r^2 e^{-\frac{r^2}{351}}+\cdots\,, &&\beta=\frac16\,,\label{full-pow-beta1/6}\\
&\vp_{pow}(r) + C\, r^4 e^{-\frac{33223}{31941}r}+\cdots\,, &&\beta=-\frac59\,,\label{powm59}
\eeal
where $p_3<0$ is given by \eqref{p3} and the ellipses stand for further subleading terms that will mix powers of the new piece and its free parameter, $C$, with the powers of the terms in \eqref{solution}. The power $p_3$ is plotted in fig. \ref{fig p3}. Since the authors of  \cite{Demmel2015b} use $\beta=1/6$, the solution \eqref{full-pow-beta1/6} is of particular interest. We show in sec. \ref{sec:numer-pow} that it provides a match to their numerical solution, as far as it was taken.

\paragraph{(f)} Except for $\beta=0$, and $\beta=\beta_\pm$ as in \eqref{beta n=2}, as discussed in sec. \ref{exceptionss}, the asymptotic series
\be 
\label{non-pow}
\vp(r) = r^2 f_{asy}\left(\log(r/A)\right)
\ee 
forms the basis for the asymptotic solutions below, where\footnote{$f_{asy}(x)$ should not be confused with the Lagrangian $f(R)$ defined via \eqref{dim vars}.}
\be 
f_{asy}(x)=k_1 x+k_2\log(x)+k_3\dfrac{\log(x)}{x}+\dfrac{k_4}{x}+k_5\dfrac{\log^2(x)}{x^2}+\cdots \,.
\label{vptotal}
\ee
For $\beta \ne -1/3$, $5/6$, the coefficient $k_1$ is derived in sec. \ref{sec:non-pow-leading} and is
given in \eqref{non-pow k1}, while the other $k_i$ are derived in sec. \ref{sec:non-pow-subleading} and are given in appendix \ref{kivalues}. As explained in sec. \ref{exceptionss}, for $\beta=-1/3$ and $5/6$ the coefficients take different values as given in \eqref{ki13} respectively \eqref{ki56}.
The arguments $x$ in the logs in \eqref{vptotal} can be replaced by $x/c$ as in \eqref{f} but as shown in sec. \ref{sec:non-pow-subleading} this is not an extra parameter and can be absorbed into the free parameter $A$ in \eqref{non-pow}. The full asymptotic solution then takes the following forms:
\beal 
f_{asy}(x) \phantom{+} &,\qquad\qquad\qquad\qquad\qquad\qquad \beta\in \left(\b_-,\b_+\right)\,,\label{non-pow-only}\\
f_{asy}(x) + &\,B e^{-\tfrac{2 }{3}\sqrt{-\tfrac{2}{h_3}}\,x^{\frac32}}\,,\qquad\qquad\quad \beta\in \left(-\infty,-\tfrac{5}{9}\right)\cup\left( -\tfrac{1}{3},\b_-\right) \cup\left(\b_+,\infty\right)\backslash\left\{0,\tfrac{5}{6}\right\}\,,\label{non-powExp32}\\
f_{asy}(x) + &\, \left\{ B \cos\left(\tfrac{2  }{3}\sqrt{\tfrac{2}{h_3}}\,x^{\frac{3}{2}}\right)+C \sin\left(\tfrac{2  }{3}\sqrt{\tfrac{2}{h_3}}\,x^{\frac{3}{2}}\right) \right\} \, e^{4h_2 x/h_3}\,,\ \
\beta\in \left(-\tfrac{5}{9},-\tfrac{1}{3}\right)\backslash\left\{-0.4111\right\}\,,
\label{non-powOsc32}\\
f_{asy}(x) + &\, B\, e^{-L_+x}\,, \qquad\qquad \qquad\qquad\beta=-\frac13\,, \label{non-powExpL}\\
f_{asy}(x) + &\,B %\exp\left\{
e^{-\tfrac{2\sqrt{21}}{15}\,x^{\frac{3}{2}}}\,,
%\right\}\,,
\qquad\qquad \qquad\quad\beta=\frac56\,,\label{non-powExp32alt}\\
f_{asy}(x) + &\,B\, e^{-\frac{23056}{22815}\,e^x}\,,\qquad\qquad \qquad\quad\beta=-\frac59\,,\label{non-powExpExp}\\
f_{asy}(x) + &\, B\, x^q \cos\left(L_I\,x^{\frac{3}{2}}\right)+C\, x^q \sin\left(L_I\,x^{\frac{3}{2}}\right)\,,\qquad\qquad\qquad \beta=-0.4111\label{non-powOsc}
\eeal
where the positive square root is taken, $h_2$ and $h_3$ are defined in \eqref{h2nonpow} and \eqref{h3nonpow},  $L_+$ in \eqref{Lpm}, $L_I=1.0648$ and $q=-2.1499$.
The top three solutions are derived in sec. \ref{missing params 2}. However \eqref{non-pow-only} required a separate analysis for $\beta=1/6$ and $0.3800$, in secs. \ref{sec:third derivative} and \ref{k2sec} respectively. The next two are derived in sec. \ref{sec:altered}, \eqref{non-powExpExp}is derived  in sec. \ref{sec:third derivative}, and \eqref{non-powOsc} is derived in sec. \ref{k2sec}.

\subsection{Physical part}
\label{sec:phys}
Using \eqref{phys}, we can now extract the corresponding physical parts. As noted in sec. \ref{sec:intro-phys}, these give the universal equation of state precisely at the fixed point, analogous to \eqref{Vscale} in a scalar field theory. We see that except for the cases (d) and (f) discussed below, the result vanishes:
\be 
f(R) |_{\text{phys}} = 0\,.
\ee
The asymptotic solution \eqref{full-pow-beta1/6} which matches the numerical solution in 
\cite{Demmel2015b}, thus also falls in this class.
Similar results were obtained for cases in the conformal truncation model of \cite{Dietz:2015owa,Dietz:2016gzg} where also divergent results were found. Perhaps these indicate that these do not give a sensible continuum limit, although a fuller understanding is needed for example by moving away from the fixed point by including relevant couplings.

For case (d), from \eqref{n=2sol} we get (for $\beta=\beta_\pm$):
\be 
f(R) |_{\text{phys}} = A R^2\,.
\ee
This equation of state was also found in ref. \cite{Dietz:2012ic} for the $f(R)$ approximation given in \cite{Benedetti:2012}, and for solutions found in ref. \cite{Ohta2016}.
For any of the solutions for case (f),  we get from \eqref{non-pow} and \eqref{vptotal} that
\be 
f(R) |_{\text{phys}} = k_1 R^2 \log(R) - R^2 \log(Ak^2)\,,
\ee
and $k_1$ is given by \eqref{non-pow k1}. Since we require $k\to0$, the second term is a positive logarithmic divergence. It is perhaps a signal of the asymptotic freedom of the $R^2$ coupling in this case where thus it should be treated as in ref. \cite{Falls:2016msz}. As we will see in the next section, global solutions with the asymptotics of case (f) exist only for $\beta<0$ and therefore $k_1$ is always positive.

\subsection{Dimensionality of the fixed point solution spaces}
\label{sec:dimensionality-ef}

Using \eqref{counting} and table \ref{table: fixed sings} we can read off the dimensionality of the corresponding fixed point solution spaces for cases (e) and (f). 

\paragraph{(e)} We see that  since \eqref{solution} has two free parameters, \eqref{p3pos} only extends to global solutions for $\beta$ in the negative interval. When $\beta>-5/9$ the fixed points, if any, form a discrete set, while lines of fixed points are found for $\beta<-5/9$. The other solutions, \eqref{p3neg} -- \eqref{powm59}, all have three free parameters and thus provide no constraints on the dimension of the solution space, which thus follows the pattern discussed in sec. \ref{sec:fixed-sing-count}. In particular \eqref{p3neg} extends to a global solution only for $\beta<1/6$ where it can have discrete fixed points or lines of fixed points depending on the sign of $\beta$, \eqref{full-pow-beta1/6} has discrete solutions, and \eqref{powm59} has planar regions of fixed points. 

\paragraph{(f)} Using \eqref{beta n=2}, we see that  since \eqref{vptotal} has only one free parameter, solutions with asymptotic behaviour \eqref{non-pow-only} and \eqref{non-powExp32alt} do not exist since they are overconstrained by the fixed singularities, \eqref{non-powExp32} has discrete solutions for $-1/3\le\beta<0$ and lines of fixed points for $\beta\le -5/9$, \eqref{non-powExpL} has discrete solutions, and 
\eqref{non-powOsc32}, \eqref{non-powExpExp} and 
\eqref{non-powOsc} all generate lines of fixed points.

\subsection{Which fixed point?}
\label{sec:which}

From the point of view of the asymptotic safety programme, it would be phenomenologically preferable if the correct answer lay in only one of the discrete sets: \eqref{p3pos} for $\beta\in(-5/9,-0.1809)$, \eqref{p3neg} for $\b\in(0,\,1/6)$, \eqref{full-pow-beta1/6} (the choice made in ref. \cite{Demmel2015b}), \eqref{non-powExp32} for $\b\in(-1/3,\,0]$, or \eqref{non-powExpL}. However we need a convincing argument for choosing one solution over the others. 

Inspection of the form of the cutoff functions used in the derivation of \eqref{fp}, see eqn. (3.13) in ref. \cite{Demmel2015b}, shows that $\beta<0$ corresponds to cases where some scalar modes never get integrated out, no matter how small we take $k$. One therefore could argue that for $\beta<0$ the Wilsonian renormalization group is undermined. Continuous sets of solutions \cite{Dietz:2012ic} were also found with another approach to the $f(R)$ approximation \cite{Benedetti:2012}, and there also there is a scalar mode that never gets integrated out. Clearly it would be very useful to know if this correlation is found for other formulations \cite{Percacci:2015wwa,Labus:2015ska,Eichhorn:2015bna,Demmel:2014fk,Demmel:2012ub,Demmel:2013myx,Benedetti:2013jk,Demmel:2014hla,Demmel2015b,Ohta:2015efa,Ohta2016,Percacci:2016arh,Falls:2016msz,Ohta:2017dsq} in the literature.

For these continuous solutions it could also be, like in ref. \cite{Dietz:2012ic}, that the $f(R)$ approximation is breaking down there, such that the whole eigenspace becomes redundant \cite{Dietz:2013sba}. To check this would require developing the full numerical solutions. 

It could also be that these are artefacts caused by violations of background independence 
\cite{Bridle:2013sra}. We saw there that at in the Local Potential Approximation, providing the modified split Ward identity that reunites the fluctuation scalar field $\phi$ with its background counterpart $\bar{\phi}$, is satisfied, the spurious behaviour is cured.  The implementation of background scale independence in refs. \cite{Morris:2016spn,Percacci:2016arh,Ohta:2017dsq} is arguably the equivalent step for the $f(R)$ approximation, since it reunites the constant background curvature $\bar{R}$ with the multiplicative constant conformal factor piece of the fluctuations. We saw there that the resulting formulations can be close to the single-metric approximation used to derive \eqref{fp},  in the sense that minimum changes are needed \eg setting space-time dimension to six, or choosing a pure cutoff, to convert the fixed point equation into a background scale independent version. It could therefore be promising to investigate formulations with these changes.

On the other hand, continuous solutions were found in the conformal truncation model of ref. \cite{Dietz:2016gzg}, where a clear cause was found in the conformal factor instability \cite{Gibbons:1978ac}.  The cutoff implementation did not introduce fixed singularities, background independence was incorporated \cite{Dietz:2015owa}, all modes were  integrated out, and an analogous breakdown to the $f(R)$ approximation \cite{Dietz:2013sba} was either not there or not possible. This suggests that the issues go deeper.

\paragraph{} In the next two sections, we provide the details of how the asymptotic solutions were discovered and developed.

\section{Asymptotic expansion of power law solutions}
\label{sec:pow-laws}

We now give the detailed study of the asymptotic behaviour of the solution $\vp(r)$ in the IR limit $r \rightarrow \infty$. (This is equivalent to the large background curvature ${R}$ limit for fixed $k$, by \eqref{dim vars}. For fixed ${R}$, $r \rightarrow \infty$ corresponds to what we more commonly refer to as the IR limit, $k\rightarrow 0$.) Finding such an asymptotic solution initially requires a degree of guesswork. A profitable place to start is to assume that the asymptotic series starts with a power, \ie
%solution can be approximated by an asymptotic series of the form
\be
\label{ansatz}
\vp(r)= A\,r^n + \cdots\,,
\ee
where  $A\ne0$ is typically an arbitrary coefficient, and subsequent terms need not be powers but are successively smaller than the leading term for large $r$. 

Now, requiring that \eqref{ansatz} satisfies the fixed point equation \eqref{fp}, means that at large $r$, the leading piece in this equation must itself satisfy the equation. In this way we typically determine $n$ and sometimes also $A$. The leading piece of the fixed point equation will be satisfied either because the left-hand side and the right-hand side  provide such a piece and for appropriate values of $n$ and $A$ these are then  equal, or because only one side of the equation has such a leading piece but this can be forced to vanish by appropriate values of $n$ and $A$. In this sense we require the leading terms to `balance' in the fixed point equation. As we will see, requiring then the sub-leading terms also to balance will determine the form of the corrections needed in \eqref{ansatz}.

\subsection{Leading behaviour}
\label{Leading behaviour}
We begin by finding the leading behaviour of $\vp(r)$  \ie solving for the power $n$ in \eqref{ansatz}. We build the asymptotic series leaving $\beta$ unspecified for the reasons given at the end of sec. \ref{sec:fixed-sing-count}.
We plug the solution ansatz \eqref{ansatz} into the fixed point equation \eqref{fp}, expand about $r=\infty$ and keep only the leading terms in the large $r$ limit. Since the coefficients \eqref{coeffs-c} are expanded along with everything else and only their leading parts are kept, it is useful to introduce the following definitions:
\begin{align}
\label{coeffs d}
 \tilde{c}_1\sim\,\tilde{c}_2 &\sim\,-\frac{5\beta(2\beta-1)(6\beta-5)}{768\pi^2}%\beta(2\beta-1)(6\beta-5)
%  -\frac{5 \left(108 \beta ^3-144 \beta ^2+45 \beta \right)}{6912 \pi ^2} 
\,r^3\,\equiv \tilde{d}_1 \, r^3 \,,\nonumber\\
%\tilde{c}_2 &\sim\, -\frac{5 \left(108 \beta ^3-144 \beta ^2+45 \beta \right)}{6912 \pi ^2} \,r^3 \,
%\equiv \tilde{d}_2 \,r^3\,, \nonumber\\
c_1 &\sim\, -\frac{\beta(6\beta-1)^2}{2304\pi^2}
%-\frac{36 \beta ^3-12 \beta ^2+\beta }{2304 \pi ^2} 
\,r^3 \,
\equiv d_1 \, r^3 \,,\nonumber\\
c_2 &\sim \, -\frac{\beta (6\beta-1)^2(9\beta+1)}{4608\pi^2}
%-\frac{324 \beta ^4-72 \beta ^3-3 \beta ^2+\beta }{4608 \pi ^2} 
\,r^4 \,
\equiv d_2\,r^4 \,,\nonumber\\
c_4 &\sim\, \frac{\beta (6\beta-1)^2(9\beta+5)}{4608\pi^2}\,
%\frac{324 \beta ^4+72 \beta ^3-51 \beta ^2+5 \beta }{4608 \pi ^2}
r^4\,
\equiv d_4\, r^4\,,
\end{align}
where we have rewritten $\tilde{c}_1$ and $\tilde{c}_2$ using $\alpha=\beta-2/3$. 
%and where $\tilde{d}_i$ and $d_i$ are functions of $\beta$ only and do not depend on $r$. 
For functions $f(r)$ and $g(r)$, $f(r) \sim g(r)$ means that $\lim_{r\rightarrow\infty}\,{f(r)}/{g(r)} = 1$. We note that for certain values of $\beta$, the leading behaviour of the coefficients will be different from those given in \eqref{coeffs d} since the leading coefficients will vanish. We discuss this in section \ref{Exceptions}, and comment there and below on the case of $\beta={1}/{6}$.

Inserting ansatz \eqref{ansatz} into the fixed point equation \eqref{fp}, we find that the leading piece on the left-hand side as $r\rightarrow\infty$ is simply
\be
\label{lhs}
(4-2n) A r^n\,,
\ee
and the leading piece on the right-hand side is given by
\be
\label{rhs}
\left\{ \frac{ n(3-2n)\, \tilde{d}_1 }{3-3  \beta  n+ n }
+\frac{ n d_1+n (n-1) d_2-2 n(n-1) (n-2)   d_4}{ n(n-1) \left(9 \beta ^2+6 \beta +1\right) -3    n\beta-2 n+2 } \right\} 
r^2\, ,
\ee
where we  substituted $\alpha=\beta-2/3$ in the first fraction. The important observation here is that the left-hand side goes like $r^n$ whereas the right-hand side goes like $r^2$, and thus which side dominates will be determined by whether $n$ is less than, greater than or equal to 2. Below we investigate these possible scenarios to determine the power $n$. We recognise that the scaling behaviour of the right-hand side could differ from $r^2$ if cancellations were to occur in either the denominators or the numerators. This is discussed in section \ref{Exceptions}.%\hfill\\

\paragraph{\textbf{Scenario (1):}}
For $n>2$, the left-hand side \eqref{lhs} dominates and thus we require that $n$ be chosen to set the left-hand side to zero.
%\be
%\label{n>2}
%4 A r^n - 2 n A r^n =0\,.
%\ee
However we see immediately that this is only true if $n=2$ and thus we reach a contradiction.
%\newline

\paragraph{\textbf{Scenario (2):}}
For $n\le2$, the leading right-hand side part, \eqref{rhs}, must vanish on its own. For $n<2$ this is because the right-hand side dominates, while for $n=2$ it must be so because we have just seen that in that case the left-side vanishes on its own. Thus for this scenario, we require $n$ to be such that the coefficient in \eqref{rhs} vanishes.
%\be
%\label{scenario2}
%0 = \frac{ n\tilde{d}_1-2 n (n-1) \tilde{d}_2 }{3-3  \beta  n+ n }
%+\frac{ n d_1\,+n (n-1) d_2\,-2 n(n-1) (n-2)   d_4 }{ n(n-1) \left(9 \beta ^2+6 \beta +1\right) -3    n\beta-2 n+2 }\, .
%\ee
Solving this we find for generic $\beta$, four solutions for $n$, which we now describe.

\paragraph{} One solution is 
\be
\label{n=3/2}
n=3/2 \,.
\ee
We will pursue this asymptotic solution in sec. \ref{Sub-leading behaviour}, where, as we will see, it allows us to build a legitimate asymptotic series.  We work this out in detail to demonstrate the general method. 
%We will also see in sec. \ref{sec:numer-pow} that it matches the numerical solution found in ref. \cite{Demmel2015b}. 
%Also at this stage,
%%for the purposes of illustration of the general method, and because, 
%depending on the nature of the correction terms, it could conceivably match the numerical behaviour found  in  ref. \cite{Demmel2015b}. 
As we will see in section \ref{Missing parameters} there are values of $\beta$ for which this choice of scaling is not allowed. However we can say that for generic of $\beta$ one possibility for the leading term in an asymptotic expansion \eqref{ansatz} is therefore
\be
\label{LO}
\vp(r) \sim A \, r^{3/2}\,.
\ee
%As well as being the simplest choice we can make for $n$ (comparing to the alternative solution with generic $\beta$ \eqref{double root soln}), the choice is further justified in section \ref{Sub-leading behaviour} where, as we will see, it allows us to build a legitimate asymptotic series. 
This leading behaviour agrees with the quantum scaling found in \cite{Demmel2015b}, but is in conflict with the classical and balanced scaling $r^2$ that the authors ultimately use to approximate the large $r$ behaviour of their solution. In fact the full solution we find, namely \eqref{full-pow-beta1/6}, %in secs. \ref{Sub-leading behaviour} and \ref{Missing parameters} 
does not agree with that suggested in ref. \cite{Demmel2015b}, 
but  can match quite acceptably the numerical solution they found, as we show in sec. \ref{sec:numer-pow}.

%turn out to rule out matching to their numerical solution, but we emphasise that \textit{a priori}
%other numerical solutions not so far investigated could indeed match into this asymptotic solution, as is also the case for the other asymptotic solutions we find.
%Again it seems unlikely that this will provide us with an asymptotic solution that matches the numerical one in  ref. \cite{Demmel2015b}, however for the purposes of illustration we will pursue this as an example in the remainder of the paper.
%As explained in section \ref{Exceptions}, when $\beta=\frac{1}{6}$ the only solution for $n$ is that of \eqref{n=3/2}.

Another solution is $n=0$. This is already clear from \eqref{fp} and follows from the fact that the numerators depend only on differentials of $\vp$. Setting $n=0$, we have arranged that the leading $r^2$ term vanishes, so now we turn to the subleading term. We see that the left-hand side of \eqref{fp} will provide an $r^0$ piece. For $\vp=A$ to be the beginning of an asymptotic series we will need to balance this piece with a term on the right-hand side. Whatever subleading term we add, it will generically no longer be annihilated by the numerators in \eqref{fp}. Meanwhile the denominators will go like a constant for large $r$ (from the undifferentiated $\vp$ parts). Thus, using \eqref{coeffs d}, we see by inspection that the subleading piece goes like $1/r^2$. By expanding \eqref{fp} in an asymptotic expansion and matching coefficients we thus find eqn. \eqref{n=0} with
\be 
\label{n=0 k1}
k_1 = \frac{18432\pi^2A^2}{7\beta(972\beta^3+528\beta^2-497\beta+117)}\,.
\ee
We could continue to investigate this asymptotic solution, developing further subleading terms and finding out how many parameters it ultimately contains, but  this solution should be a very poor fit to the numerical solution found in ref. \cite{Demmel2015b} which numerically shows  behaviour identified  in ref. \cite{Demmel2015b} as $\vp\sim r^2$ for large $r$.

\begin{figure}
\centering
\includegraphics[scale=1.3]{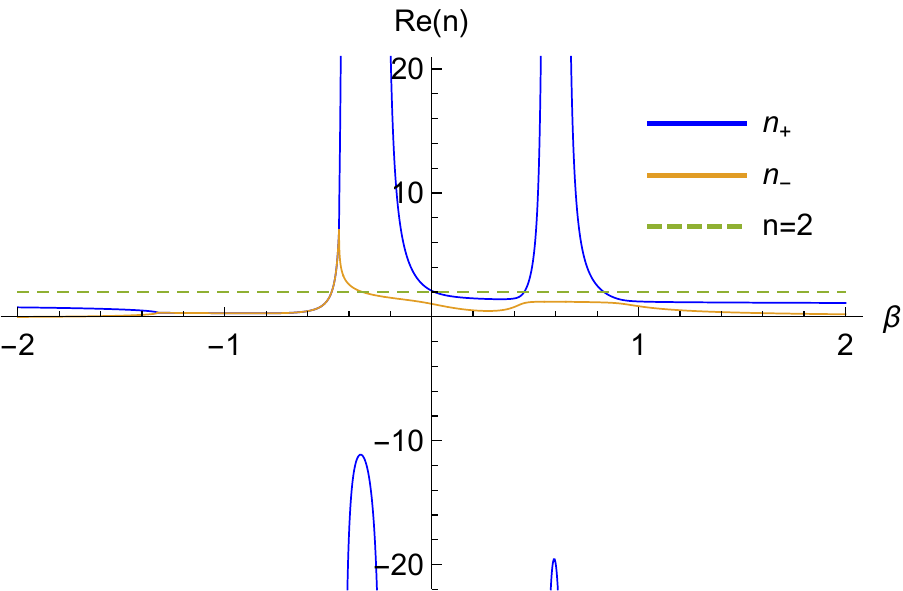}
\caption{Plot of the two solutions $n_\pm$ given by \eqref{double root soln}.} 
%$n_+$ and $n_-$ label the solutions with positive and negative square roots respectively.}
\label{scenario 2}
\end{figure}
The last two solutions are functions of $\beta$ and are given by the roots of the quadratic
\begin{multline}
\label{double root soln}
0=\left( 4212\,{\beta}^{4}-2268\,{\beta}^{3}-1395\,{\beta}^{2}+486\,
\beta+145 \right) {n}^{2}\\
+ \left( -4212\,{\beta}^{4}+2043\,{\beta}^{2}
+147\,\beta-458 \right) n
+972\,{\beta}^{3}+1152\,{\beta}^{2}-1185\,
\beta+321\,.
\end{multline}
%\begin{align}
%\label{double root soln}
%n=& \,\frac{458+3 \beta  \left(1404 \beta ^3-681 \beta -49\right)}{2 \left(4212 \beta ^4-2268 \beta ^3-1395 \beta ^2+486 \beta +145\right)}\nonumber\\
%&\pm\frac{1}{2 \left(4212 \beta ^4-2268 \beta ^3-1395 \beta ^2+486 \beta +145\right)}\nonumber\\
%&\quad\times \bigg(17740944 \beta ^8-16376256 \beta ^7-27801144 \beta ^6+34601256 \beta ^5\nonumber\\
%&\qquad\quad -3587895 \beta ^4-5902794 \beta ^3+1576881 \beta ^2-71376 \beta +23584\bigg)^{1/2}.
%\end{align}
The solutions \eqref{double root soln} are plotted in fig. \ref{scenario 2}.
These roots take complex values when $-1.326<\b<-0.4474$ and so the allowed solutions which we want are  those for which Re$(n)\le2$. The only region where there is not a solution Re$(n)\le2$ is from the point where Re$(n)$  crosses the $n=2$ line in this range, namely at $\b=-0.4835$, through to the point where the `blue' root diverges, namely $\b=-0.4273$.
%As we will see below, $n=2$ is however excluded. 

 In order to know whether the Re$(n)\le2$ solutions of \eqref{double root soln} 
 really lead to valid asymptotic series we need to take the expansion to the next order and check that the next order is genuinely sub-leading. We  pursue this in a similar way to the $n=0$ case above. We know that the left-hand side of \eqref{fp} $\sim r^n$ and this term will need balancing by terms on the right-hand side. On the right-hand side the denominators also $\sim r^n$, whereas the numerators, which would have gone like $\sim r^{n+2}$ have had the corresponding coefficient cancelled by choosing \eqref{double root soln}. Therefore the subleading term we need to add is a piece $\zeta(r)$ such that 
\be 
r^3 \zeta'(r) /r^n \sim r^n\,,
\ee
where we have trialled just the first term in the first numerator on the right-hand side of \eqref{fp} and compared it to the left-hand side. Solving this gives $\zeta\sim r^{2n-2}$. Since $2n-2 = n + (n-2)$, we see that this is genuinely subleading only if Re$(n)<2$. 
%(\ie excluding the $n=2$ case). 
By inspection, such a power law solution then works out for the full numerators on the right-hand side of \eqref{fp}. Substituting these first two terms of the fledgling asymptotic series into \eqref{fp} and expanding for large $r$ we find the coefficient and thus confirm our last two power-law asymptotic  solutions \eqref{n general subleading} where
%exist only in the regions of $\beta$ where Re$(n)<2$, and
\be
\label{n general k1}
k_1 = \frac{6912{A}^{2}{\pi }^{2} ( n-1 ) ^{2}{n}^{3}}{36{n}^{2} ( 16n-19)  ( n-1) ^{2}{\beta}^{2}-6n ( n-1 )  ( 76{n}^{3}-196{n}^{2}+140n-15) \beta+ c(n) }\,,
\ee
where we have set
\be
c(n) = ( n-3 )  ( 36{n}^{4}-139{n}^{3}-
210{n}^{2}+713n-420)\,.
\ee
When \eqref{double root soln} has real roots, those with $n<2$ are taken. When  complex, the real part of \eqref{n general subleading} should be taken leading to $\vp(r) \sim r^{\text{Re}(n)} \sin(\text{Im}(n)\log r +B)$ type behaviour.

For $n=2$, both the left and right-hand side scale as $r^2$ and therefore could be expected to balance. However, as we have already seen, if $n=2$ the left-hand side is identically zero and so again we require the right-hand side \eqref{rhs} to vanish, but now with $n$ fixed to $n=2$.  
As we will see in sec. \ref{sec:non-pow} this impasse gives a clue however to a non-power law asymptotic solution. 
Pursuing for now the power law case \eqref{ansatz} but with fixed $n=2$, we find that this equation is satisfied either for $\beta=\beta_\pm$ where $\beta_\pm$ is defined in \eqref{beta n=2},
or apparently for $\beta=0$, since then all the $\tilde{d}_i$ and $d_i$ vanish. The exceptional case of $\beta=0$ is discussed in section \ref{Exceptions}. The solutions \eqref{beta n=2} are just values of $\beta$ such that one of the roots of \eqref{double root soln} is indeed $n=2$. Substituting $n=2$ in \eqref{beta n=2} gives a quartic in $\b$, but the other two roots, $\b=-1/3$, $5/6$, are cancelled at $n=2$ by the denominator in \eqref{rhs}. The remaining quadratic is in fact the one that appears in the denominator of coefficients \eqref{h2nonpow} and \eqref{h3nonpow} that we will come across later.

To demonstrate the validity of the $n=2$ solution \eqref{beta n=2}, we need to show that the expansion can be taken to the next order. We know the expansion for general $n$ given in \eqref{n general subleading} breaks down for $n=2$. In fact in this case, since the left-hand side of \eqref{fp} already vanishes, the subleading term comes from the next term on the right-hand side in a large $r$ expansion. In this way we see that the asymptotic solution is \eqref{n=2sol} where 
\be 
\label{n=2sol k1}
k_1={\frac {312\,A \left( -21353+363048\,\beta \right) }{17166809088\,A{
\pi }^{2}\beta-14017536\,A{\pi }^{2}+10800590\,\beta-110555}}\,,
\ee
and $\beta$ is either root in \eqref{beta n=2}.

We see that overall there are in general four types of power-law asymptotic solutions given by the power being one of the roots \eqref{double root soln} providing $\beta$ is such that Re$(n)\le$ 2, or two $\beta$ independent cases: $n=0$ and $n=3/2$.

\subsection{Exceptions}
\label{Exceptions}
In general,
as mentioned previously, we recognise that certain choices for $\beta$ alter the scaling behaviour of the right-hand side of the fixed point equation such that it differs from $r^2$ in the limit $r\to\infty$. 

\subsubsection{Exceptions from the denominators}
\label{excep-denom}
For instance, the leading behaviour could increase to $r^3$ if the $r^n$ terms in one or both of the denominators were to cancel amongst themselves. These cancellations occur in the first and second fractions respectively when
\be
\label{except1}
n=\frac{3}{3\beta-1}
\ee
or
\be
\label{except2}
n=\frac{3+9 \beta +9 \beta ^2\pm\sqrt{81 \beta ^4+162 \beta ^3+63 \beta ^2+6 \beta +1}}{2 \left(9 \beta ^2+6 \beta +1\right)}\, .
\ee
%The leading power on the right-hand side could further increase if then the $r^{n-1}$ terms were to also cancel and so on. 
We will concentrate on the asymptotic solution with $n=3/2$.
For this value, \eqref{except1} and \eqref{except2} are satisfied when $\b=1$ and $\b=\pm{1}/{\sqrt{27}}$ respectively. For these values of $\b$, a leading power of $n={3}/{2}$ is not allowed and instead to find the leading behaviour in these cases we must treat $\b=1$ and $\b=\pm{1}/{\sqrt{27}}$ separately from the start. As we will see in these cases we just recover the other three power-law solutions. 
%As we discuss below, exceptions can also occur at special values of $\beta$ when the leading power in the numerator is reduced. We now list all of these exceptions.

%\paragraph{$\b=1$:}
\paragraph{${\boldsymbol{\b=1}}$:} the leading piece in the limit $r\rightarrow\infty$ on the left-hand side is still of course given by \eqref{lhs}. 
%Indeed this will be the case for any $\beta$ as the left-hand side contains no coefficients $c_i,\tilde{c}_i$ and is therefore independent of $\beta$. 
Also, since $\beta=1$ does not correspond to one of the values at which the leading parts of the coefficients $c_i,\tilde{c}_i$ vanish, see \eqref{coeffs d}, the leading piece on the right-hand side will still be given by \eqref{rhs}, but now with $\beta$ set to 1:
\begin{equation}
\label{rhs beta 1}
-\frac{5 n r^2}{768 \pi ^2} - \frac{25 \left(14 n^3-37 n^2+24 n\right) r^2}{2304\pi ^2 \left(16 n^2-21 n+2\right)}\,.
\end{equation}
Note that even though $\beta=1$ has been identified as a value at which the $r^n$ terms in the first denominator of \eqref{rhs} vanish, this is only when $n=\frac{3}{2}$ and so here we still see the right-hand side scaling as $r^2$. The scaling of the right and left-hand sides is the same as that in section \ref{Leading behaviour} and so by the same reasoning we see that $n$ must be less than 2 and therefore \eqref{rhs beta 1} must vanish in order to satisfy the fixed point equation for large $r$. %($n$ cannot equal 2 as this requires $\beta=\frac{1}{78}(18+\sqrt{285})$, see scenario 2, section \ref{Leading behaviour}).
The right-hand side \eqref{rhs beta 1} vanishes when $n=(62 \pm\sqrt{127})/59$ and also trivially for $n=0$. Indeed these are the remaining power-law solutions, in particular the former pair are the values for $n$ given by \eqref{double root soln} with $\beta=1$ as expected, while the latter is the solution \eqref{n=0}.

%\paragraph{$\b=\pm\frac{1}{\sqrt{27}}$\newline}

\paragraph{$\mathbf{\boldsymbol{\beta=}\pm{1}/{\sqrt{27}}}$:} the right-hand side of the fixed point equation again scales like $r^2$ in the large $r$ limit, as this $\beta$ is also not one of the exceptional values appearing in \eqref{coeffs d}. 
%\be
%-\frac{5 \left(49 \sqrt{3}\mp48\right) n (2 n-3) r^2}{20736 \pi ^2 \left(\left(\sqrt{3}\mp3\right) n\mp9\right)}
%-\frac{n \left(\left(23 \sqrt{3}\mp39\right) n-37 \sqrt{3}\pm63\right) r^2}{41472 \pi ^2 \left(\left(\sqrt{3}\pm2\right) \mp2\right)}\,.
%\ee
We see that again $n$ must be less than 2 and that the right-hand side must vanish. Once again, the values of $n$ just correspond to  \eqref{double root soln}  when $\beta=\pm{1}/{\sqrt{27}}$.
% we have
%\be
%n=\frac{-441+3493 \sqrt{3}\pm\sqrt{18391764-6570882 \sqrt{3}}}{4 \left(603+446 \sqrt{3}\right)}
%\ee
%and when $\beta=-\frac{1}{\sqrt{27}}$, we have
%\be
%n=\frac{441+3493 \sqrt{3}\pm\sqrt{18391764+6570882 \sqrt{3}}}{4 \left(446 \sqrt{3}-603\right)}\,.
%\ee
And we still also have the $n=0$ solution \eqref{n=0}.

\subsubsection{Exceptions from the numerators}
\label{sec:excep-numer}
Exceptions to the leading behaviour $n=\nfrac{3}{2}$ also arise from particular choices for $\beta$ reducing the powers of $r$ appearing in the coefficients $\tilde{c}_i$ and $c_i$ which could result in an overall decrease in the leading power on the right-hand side of the fixed point equation. The values of $\beta$ for which the leading power of $r$ in the coefficients vanishes can readily be read from \eqref{coeffs d}.
%are listed in table 2 alongside the corresponding affected coefficient.
%\begin{table}
%\label{coeffs c}
%\begin{center}
%\begin{tabular}{ c|c }
%Coefficient & $\beta$\\
%\hline
%$\tilde{c}_1$ & $0,\, \frac{1}{2},\, \frac{5}{6}$ \\ 
%$\tilde{c}_2$ & $0,\, \frac{1}{2},\, \frac{5}{6}$\\ 
%$c_1$ & $0,\, \frac{1}{6}$ \\ 
%$c_2$ & $-\frac{1}{9}\,,0\,,\frac{1}{6}$\\
%$c_4$ & $-\frac{5}{9}\,,0\,,\frac{1}{6}$\\
%\end{tabular}
%\end{center}
%\caption{Coefficients $\tilde{c}_i$ and $c_i$ alongside the values of $\beta$ for which their leading parts vanish.}
%\end{table}
Notably however, both numerators in \eqref{rhs} are satisfied independently for $n=\nfrac{3}{2}$. This means that for the values $\beta=\nfrac{1}{2},\nfrac{5}{6},\nfrac{1}{6}$ for which only one of the fractions becomes sub-dominant the $\beta$-independent solution $n=\nfrac{3}{2}$ remains valid. 

\paragraph{$\boldsymbol{\beta=1/6}$:} although we are concentrating on the $n=\nfrac32$ solution, for completeness we note that $\beta=\nfrac16$ does present an exception for the general power solutions \eqref{double root soln}. From \eqref{double root soln}, we would expect to find asymptotic series with leading powers $n= (19\pm\sqrt{73})/18 = 1.530, 0.5809$. However when $\beta=\nfrac16$, we see from \eqref{coeffs d} that $d_1$, $d_2$ and $d_4$ all vanish. Then from \eqref{rhs} we see that in this case the only solutions left for $n$ are the $n=0$ and $n=\nfrac32$ cases established in sec. \ref{Leading behaviour}.

\paragraph{$\boldsymbol{\beta=0}$:} in this case the leading powers of $r$ in all the coefficients vanish, \cf \eqref{coeffs d}.
%For $\beta=0$ however the leading parts of all the coefficients vanish and \eqref{rhs} can no longer be considered as a valid starting point in determining the leading behaviour of the solution. 
(We see the implications of having $\b=0$ in section \ref{Sub-leading behaviour} where it represents a pole of all but one of the coefficients in the asymptotic expansion given in appendix \ref{AppendixA}.)
%\paragraph{} %$\b=0$\newline}
%\label{excpt beta=0}
As a result, the right-hand side of the fixed point equation \eqref{fp} in general no longer scales as $r^2$ but instead goes like $r$. This apparently implies two solutions: either the asymptotic series is $\vp = A r^2 + k_1 r +\cdots$, since $r^2$ satisfies the left-hand side on its own, or  the asymptotic series takes the form $\vp = Ar+\cdots$, with the $r$ term then balancing both sides of the equation. However substituting $\vp=Ar^n$ into the right-hand side of \eqref{fp} (with $\beta=0$) we find the leading term is:
\be 
\label{beta0leading}
-{\frac {5\,n \left( 58\,{n}^{3}-389\,{n}^{2}+792\,n-477 \right) }{
4608\,{\pi }^{2} \left( n+3 \right)  \left( n-1 \right)  \left( n-2
 \right) }}\, r\,,
\ee
which thus presents an exception for both of these cases! The reason for this is that $n=1,2$ happen to be precisely the two powers that reduce the leading power in second denominator in this case, as can be seen from \eqref{except2}. Thus actually when $n=1$ or $n=2$, the second term on the right-hand side of \eqref{fp} contributes $\sim r^2$. Since it does so now with no free parameters ($\beta$ and $n$ having been fixed), neither suggested asymptotic solution will work: for $n=1$ because the correction is larger than the supposed leading term, while for $n=2$ there is nothing to balance it since the left-hand side vanishes identically. Finally, we consider general $n$. For $n>1$, the left-hand side dominates and we require that it vanishes for the fixed point equation to be satisfied, but this only gives us the already excluded $n=2$ solution. For Re$(n)<1$ the right hand side dominates and \eqref{beta0leading} must vanish on its own. The cubic in the numerator has no roots in this region and thus we are left with only an $n=0$ solution, namely \eqref{n=0no2}.
Note that this differs from \eqref{n=0}, in particular the first subleading power is now $1/r$.

\paragraph{} 
To summarise for the $n=3/2$ asymptotic series in particular, we have seen that this fails to exist at $\beta = 0$, 1, and $\pm 1/\sqrt{27}$. However as we will see, in general these exceptions do not obstruct the construction of the subleading terms or the subsequent determination of the missing parameters.

\subsection{Sub-leading behaviour}
\label{Sub-leading behaviour}
In this section we present the method for determining subsequent terms in the asymptotic series \eqref{ansatz}. As we stated, we will concentrate on the $n=3/2$ solution.
We have seen in the previous section that for generic $\beta$ the fixed point equation scales like $r^2$ for large $r$. Choosing $\vp\sim A\, r^{\nfrac{3}{2}}$ causes the $r^2$ contribution to disappear, therefore satisfying the equation in the large $r$ limit. This is precisely because a leading power of $n=\nfrac{3}{2}$ is what is required to make the coefficient multiplying the $r^2$ term be identically zero.

Once the $r^2$ terms have vanished, the new leading large $r$ behaviour of the fixed point equation is $r^{\nfrac{3}{2}}$, coming from the undifferentiated $\vp$ on the left-hand side of \eqref{fp}. The next term in the solution should be such that it now cancels the pieces contributing to the new leading behaviour. It is with this in mind that we proceed to build the sub-leading terms of the solution.

We denote the next term in the solution by a function $\zeta(r)$ such that
\be
\vp(r)= A\, r^{3/2} + \zeta(r) +\cdots\,,
\ee
 where $\zeta$ grows more slowly than the leading term.
 This implies that we can if necessary find its leading corrections algorithmically, by taking large enough $r$ to allow linearising the fixed point equation in $\zeta$.
This will give us a linear differential equation for $\zeta$, where we keep only the leading parts in a large $r$ expansion of its coefficients.  This equation is set equal to the new leading piece, namely the $r^{\nfrac{3}{2}}$ piece discussed above, and is straightforward to solve since we only want the leading part of the particular integral.
In fact
inspection of the fixed point equation shows that this contribution can only come from the right-hand side and that it requires $\zeta(r)\propto r$. 
Indeed with this choice, the leading contribution from the numerators on the right-hand side are then terms which scale like $r^3$. The leading terms from the denominators scale like $r^{\nfrac{3}{2}}$ (providing no exceptional cases arise). Together these give an overall contribution of $r^{\nfrac{3}{2}}$ to the right-hand side as required. Including the next term in the solution we now have
\be
\vp(r)= A\, r^{3/2} + k_1 r +\cdots \,.
\ee
To find  the coefficient $k_1$, we substitute the solution as given above into the fixed point equation and take the large $r$ limit. Collecting all terms on one side of the equation, the leading terms go like $r^{\nfrac{3}{2}}$ as expected, multiplied by a coefficient containing $k_1$. We require this coefficient to vanish in order to satisfy the fixed point equation and so $k_1$ must take the following form
\be
k_1 = \frac{3456 \pi ^2 A^2 (\beta -1) \left(27 \beta ^2-1\right)}{\beta  \left(1620 \beta ^4-2376 \beta ^3+903 \beta ^2+2 \beta -19\right)}\,.\label{Fval}
\ee
The next terms in the series are found by repeating this procedure.
After five iterations the solution becomes \eqref{solution}
where $k_1$ is given in \eqref{Fval} and the more lengthy expressions for the other coefficients are given in appendix \ref{AppendixA}. Note that a second constant $b$, independent of $A$, is found as a result, through particular integrals containing logs. At this point our solution therefore contains two independent parameters in total.

\subsection{Exceptions}
\label{exceptions}
The solution \eqref{solution} will break down at values of $\beta$ corresponding to poles of the coefficients $k_i$. As can be seen straight-away, $\beta=0$ is one such value. This has already been flagged-up as problematic in section \ref{Exceptions} where the trouble was traced back to the fact that when $\beta=0$, the leading part of each of the coefficients in the fixed point equation \eqref{coeffs d} vanishes. This means that \eqref{rhs} no longer represents the true asymptotic scaling behaviour of the fixed point equation and should not be used to derive the leading behaviour of the solution.

There are other poles in the coefficients besides at $\beta=0$, as can be readily seen from the full form of the coefficients given in appendix \ref{AppendixA}. We find that as we build the asymptotic series, new coefficients contain new poles, not featured in earlier terms. In this way we will build a countable, but apparently infinite, set of exceptional values of $\beta$. %\textit{A priori} this set could even be infinite. It is however 
It is not clear how these exceptional values are distributed (for example whether they lie within some bounded region or not), but
since the real numbers are uncountable, we are always guaranteed values of 
%It seems that if we move far enough down the series, eventually there will not be any values of $\beta$ left for which the series does not develop a pole.
%However we are saved by Cantor's diagonal argument which can be used to prove that the set of irreducible fractions (here the coefficients) is smaller than the set of real numbers.
%No matter how many coefficients are added to the series, giving rise to new poles along the way, we can always find a real number 
$\beta$ for which there are no poles present in the series.

If we do happen to choose a value for $\beta$ that gives rise to a pole then this signals that the term in the solution containing the pole does not have the correct scaling behaviour in order to satisfy the fixed point equation in this instance.
Take the first coefficient $k_1$ as an example. At a pole of $k_1$ (all except $\beta=0$), adding a piece on to the solution that goes like $r$ does not result in an $r^{\nfrac{3}{2}}$ contribution on the right-hand side of the fixed point equation as required, because the coefficient automatically vanishes in this case.

Instead we must look for a different sub-leading term. A less simple choice but one that works nonetheless is $r\log(r)$. The reason for this is that if all the derivatives hit the $r$ factor and not the $\log(r)$ factor then again the $r^{\nfrac{3}{2}}$ piece on the right-hand side must vanish identically, since it is as though the $\log(r)$ is just a constant multiplier for these pieces. We therefore know that in the asymptotic expansion at this order, the only terms that survive have the $\log(r)$ term differentiated. But this then maps $r\log(r)\mapsto1$ which is the power-law dependence we desired for the differentiated term. We will apply the same strategy in sec. \ref{sec:non-pow}.

Taking this as the sub-leading term, such that the solution now goes $\vp(r)= Ar^{\nfrac{3}{2}} + k_1r\log(r)$, gives rise to the desired $r^{\nfrac{3}{2}}$ term on the right-hand side, but now without the same pole (\ie a different $k_1$) and we can continue to build the series solution from there. This suggests that for a solution with leading behaviour $r^{\nfrac{3}{2}}$, each new set of poles associated with a new coefficient gives rise to further appearances of $\log(r)$ in that sub-leading term and therefore a plethora of different possible solutions dependent on these exceptional values for $\beta$.

\subsection{Finding the missing parameters}
\label{Missing parameters}
The asymptotic solution \eqref{solution} contains only two parameters, and yet we are solving a third order ordinary differential equation. Local to a generic value of $r$ we know there is a three parameter set of solutions. 
In this section we linearise about the leading solution \eqref{solution} to uncover the missing terms  \cite{Morris:1994ki,Morris:1994ie,Morris:1994jc,Dietz:2012ic,Dietz:2016gzg}. We do so by writing $A \mapsto A + \epsilon\eta(r)$ such that the solution becomes
\be
\label{soln change A}
\vp(r) = (A+\epsilon\eta(r))\, r^{\frac{3}{2}} + \cdots\,,
\ee
where $\epsilon\ll1$ and $\eta$ is some arbitrary function of $r$. Since the constant $A$ introduced in \eqref{ansatz} can take any value, we are permitted to change it by any constant amount. %and the corresponding $\vp(r)$ given by \eqref{soln change A} should still be a solution to the fixed point equation \eqref{fp} up to $\mathcal{O}(\epsilon)$ 
Thus a constant $\eta(r)$ should be a solution
in the asymptotic limit $r\rightarrow \infty$. We use this reasoning to help find the missing parameters. In sec. \ref{sec:non-pow} we will follow a related but different strategy.

We insert \eqref{soln change A}, complete with all modified sub-leading terms, into the fixed point equation \eqref{fp} and expand about $\epsilon=0$. We know already that at $\mathcal{O}(\epsilon^0)$ the fixed point equation is satisfied for large $r$, since at this order \eqref{soln change A} is equivalent to the original solution \eqref{solution}.
At $\mathcal{O}(\epsilon)$ in the large $r$ limit we obtain a third order ODE for $\eta(r)$ :
\be
\label{diff eta}
h_3\, r^5 \,\eta'''(r) + h_2\, r^4\,\eta''(r) + h_1\,r^3\eta'(r)=0\,,
\ee
where
\begin{align}
\label{coeffs f}
h_1&=\frac{5 \beta \left(-6156 \beta ^4+7020 \beta ^3-699 \beta ^2-508 \beta +83\right)}{6912 \pi ^2 A (\beta -1) \left(27 \beta ^2-1\right)}\,,\nonumber\\
h_2&= \frac{\beta  \left(-3240 \beta ^4+3078 \beta ^3+471 \beta ^2-421 \beta +47\right)}{864 \pi ^2 A (\beta -1) \left(27 \beta ^2-1\right)}\,,\nonumber\\
h_3 &= -\frac{\beta  (9 \beta +5) \left(6 \beta-1\right)^2}{576\pi ^2 A \left(27 \beta ^2-1\right)}\,.
\end{align}
Initially we would expect to have another term on the left-hand side of \eqref{diff eta} that looks like $h_0 \,\eta(r)$ times $r$ to some power. However the coefficient $h_0$ vanishes up to the order of approximation of the solution we are working to. In fact this had to be so, since otherwise $\eta(r) =\text{constant}$ would not satisfy the equation.
%his is to be expected since the shift $A \mapsto A + \epsilon\eta(r)$ for $\eta(r)$ arbitrary must also hold for $\eta(r) =\text{constant}$ and if an $h_0 \eta(r)$ term were to appear in \eqref{diff eta}, this could only be realised if $h_0=0$.
This means that what would have been be a third order ODE is instead a second order equation in $\eta'$.
This idea is analogous to the Wronskian method for  differential equations. The solution $\vp$ contains two independent parameters, $A$ and $b$, and so we already know two independent solutions of fixed point equation. These solutions can be used to build a Wronskian that satisfies a first order differential equation and which can then be used to find the unknown solution. We will not need this full machinery however.
%to show that the third unknown solution obeys a second order differential equation.

The differential equation \eqref{diff eta} is invariant under changes of scale $r\mapsto s r$ and thus has power law solutions. Setting $\eta\propto r^{p}$ we find three solutions for $p$: the trivial solution $p_1=0$ as required for consistency with the possibility of $\eta=\text{constant}$, $p_2=-\frac{3}{2}$ and
\be
\label{p3}
p_3=\frac{-4212 \beta ^4+5508 \beta ^3-1437 \beta ^2-172 \beta +53}{6 (6 \beta -1)^2 \left(\beta-1\right)\left(9\beta+5\right)}\,.
\ee
The complete solution for $\eta$  is then given by a linear combination of these powers:
\be
\eta(r) = \delta A + \delta B \,r^{-\frac{3}{2}} + \delta C\, r^{p_3}\,,
\ee
where we have introduced infinitesimal parameters $\delta A$, $\delta B$ and $\delta C$. 
Finally, inserting $\eta$ back into the solution \eqref{soln change A} we find the change in the asymptotic series complete with the change in the missing parameter
\be
\label{delta-pow}
\delta \varphi(r) \sim \delta A\, r^{\frac{3}{2}} + \delta B + \delta C\, r^{p_3 +\frac{3}{2}} \,.
\ee
The first parameter $\delta A$ resulted from perturbing the constant $A$. We see that the solution $p_2=-\nfrac{3}{2}$ was to be expected since $\delta B$ corresponds to perturbing the $b$ parameter. We have uncovered one new parameter in this case, through $\delta C$.

Whether or not the $\delta C$ perturbation is kept in the series depends on the size of $p_3$: if $p_3>0$ then the perturbation grows faster than the leading series, invalidating it,  and thus will be excluded from the solution. If $p_3< 0$ it is kept, which happens when $\beta\in(-0.1809,0.1931)\cup(0.4042,0.8913)$. However then we notice that by increasing $r$, the $r^{p_3 +\nfrac{3}{2}}$ can be made arbitrarily smaller than the leading term $Ar^{\nfrac{3}{2}}$. Therefore the full asymptotic series is developed by adding 
\be 
C\, r^{p_3 +\frac{3}{2}}
\ee
to \eqref{solution}, \ie with a now arbitrary size constant $C$. There are subleading terms to this which will look similar to those in \eqref{solution} but with an $r^{p_3}$ factor. As we develop the asymptotic series further, we will also find terms containing powers of $r^{p_3}$ coming from the non-linearity of the fixed point equation \eqref{fp}. This development  is similar to the development of asymptotic expansions in ref. \cite{Dietz:2016gzg}, where sinusoidal and log terms are also involved,  and in ref. \cite{Dietz:2012ic} where also special powers arise. 
In the present case we see that the asymptotic series takes the form of a triple expansion in $1/r$, $\log(r)$ and $r^{p_3}$ for large $r$.
The value of $p_3<0$ will determine the relative importance of all these terms. 

\begin{figure}[htbp]
\centering
\includegraphics[scale=0.7]{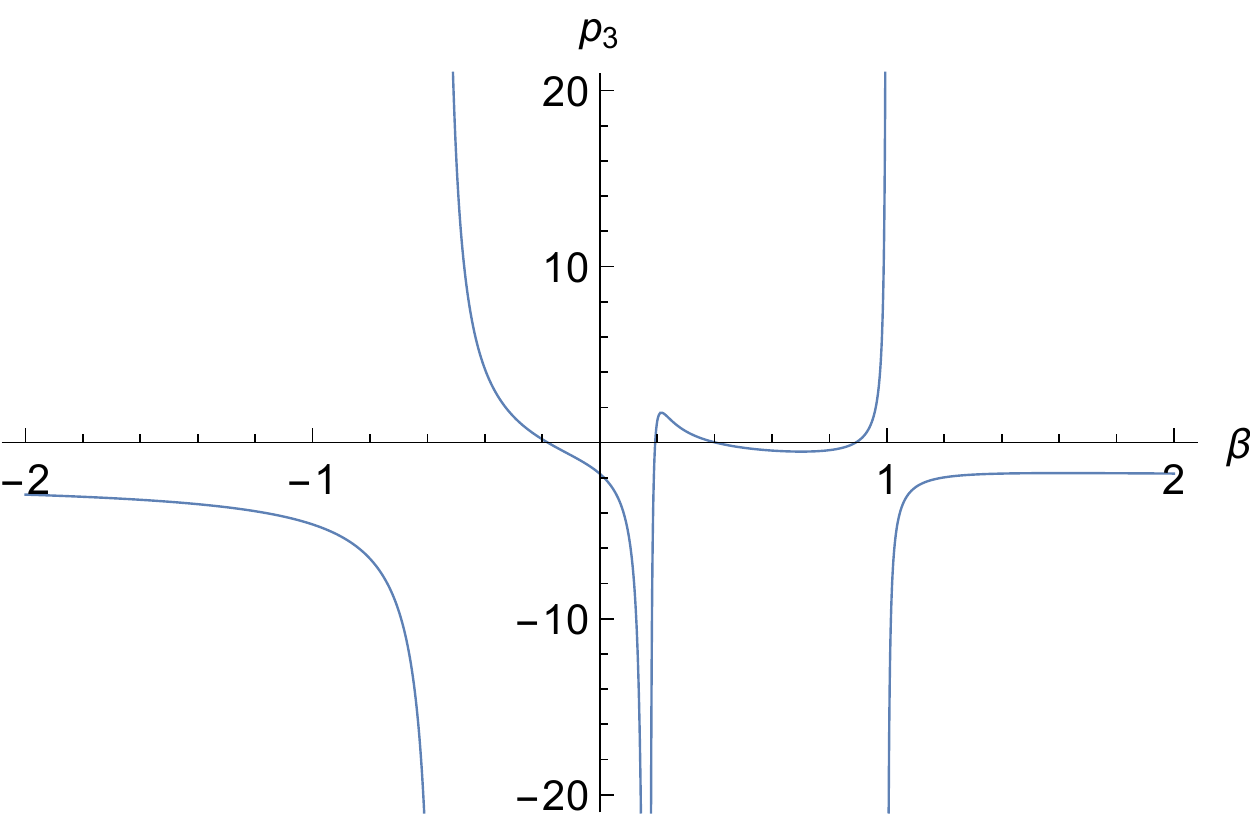}
\caption{A plot of $p_3$ against $\beta$. $p_3=0$ marks the line above which the perturbation $\delta C$ grows more quickly than the leading solution.}
\label{fig p3}
\end{figure}

The case $p_3=0$ needs careful examination: it corresponds a solution $\eta'\propto 1/r$ in \eqref{diff eta}. Therefore in this case the last term in \eqref{delta-pow} actually appears as $\delta C\, r^{\nfrac32}\log(r)$ which rules it out, since this grows faster than the leading term.
%Apparently, for $p_3=0$, the $\delta C$ perturbation is equivalent to $\delta A$ and so in this case we still have only 2 parameters in total. 
%
%An example of exclusion is seen \TRM{in the LPA for scalar field theory} where perturbations corresponding to the missing parameters grow like the exponential of a power of $\vp$, much faster than the leading solution \cite{Morris:1994ki}.
The behaviour of $p_3$ is shown in fig. \ref{fig p3}.
Knowing whether of not a missing parameter is excluded is crucial as the balance between the number of parameters and the number of constraints has important consequences for the nature of fixed point solutions.

There are three values of $\beta$ at which $p_3$ develops a pole, $\beta=-\nfrac{5}{9},\nfrac{1}{6}, 1$, as can be seen from \eqref{p3}, see also fig. \ref{fig p3}. The first two of these correspond to zeros of the coefficient $h_3$ meaning that at these values the differential equation \eqref{diff eta} is no longer the correct one and we must go to the next order in the large $r$ expansion of the $\eta'''$ coefficient. Doing this for $\beta=\nfrac{1}{6}$, we obtain the alternative equation for $\eta$:
\be
\label{diff eta 1/6}
\frac{13\,r^3}{48 A \pi^2}  \eta'''(r)+ \frac{r^4}{648 A \pi^2}  \eta''(r) + \frac{5 \,r^3 }{1296 A\pi^2}\eta'(r) = 0\,,
\ee
where the second and third coefficients are just $h_2$ and $h_1$ respectively with $\beta=\nfrac{1}{6}$ but where now the first coefficient is not the $\beta=\nfrac{1}{6}$ equivalent of $h_3$. Again there is no $\eta$ term because of the arguments given below \eqref{coeffs f} and this enables us to turn \eqref{diff eta 1/6} into a second order ODE by writing $\eta'(r)=\rho(r)$. We can then solve for $\rho(r)$. Integrating up to get $\eta(r)$ we find that for large $r$
\be
\eta(r)= \delta A + \delta B\, r^{-\frac{3}{2}} + \delta C \sqrt{r}\, e^{-\frac{r^2}{351}}\,,
\ee
%where $\delta B$ and $\delta C$ represent the missing parameters. 
Substituting $\eta$ back into \eqref{soln change A} we find
\be
\label{phi beta 1/6}
\delta \vp(r) \sim \delta A r^{\frac{3}{2}} + \delta B + \delta C\, r^2 e^{-\frac{r^2}{351}}\,.
\ee
Again, $\delta B$ results from perturbing the constant term, but now  $\delta C$  comes paired with an exponentially decaying piece. Since the exponential decays more rapidly than any power, the $\delta C$ term grows more slowly than any term we have found so far in the asymptotic solution \eqref{solution}. Therefore again we can replace $\delta C$ by $C$ and add this to the solution \eqref{solution}. Again the full series will involve powers of this term together with powers of $r$ and $\log(r)$, however even just the linear term will always be less important than any term in \eqref{solution} for sufficiently large $r$, and thus in practice one only need keep this linear term. In conclusion, when $\beta=\nfrac{1}{6}$,  the solution contains three independent parameters, $A, B$ and $C$, and takes the form of \eqref{full-pow-beta1/6}
%\be 
%\label{full-pow-beta1/6}
%\vp(r) = A\,r^{3/2} + k_1 \, r + k_2 \, r^{1/2} + k_3\, \log\left(\frac{r}{b}\right)
%+k_4\frac{\log\left(\frac{r}{b}\right)}{\sqrt{r}}+ \frac{k_5}{\sqrt{r}} +C\, r^2 e^{-\frac{r^2}{351}}+\cdots\,.
%\ee
%will always be less important than any term in \eqref{solution} and thus in practice one only need keep this linear term.
%
%the leading solution. It is reasonable to believe therefore that the last term in \eqref{phi beta 1/6} is indeed a valid part of the asymptotic expansion, albeit appearing very far down the series. 
%In conclusion, when $\beta=\frac{1}{6}$,  the solution contains three independent parameters: $A, B$ and $C$.

Following the same procedure for $\beta=-\nfrac{5}{9}$, the differential equation for $\eta$ is given by
\be
\frac{845\, r^4}{38016 \pi ^2 A} \eta'''(r) +\frac{166115\, r^4}{7185024 \pi ^2 A}\eta''(r) + \frac{830575\,r^3}{14370048 \pi ^2 A}\eta'(r) = 0\,.
\ee
%and has solution
%\be
%\eta(r)=\delta A+\delta B \,r^{-\frac{3}{2}}+\delta C \, r^{\frac{5}{2}}e^{-\frac{33223}{31941}r}\,.
%\ee
Upon substituting the solution $\eta$ back into $\vp$ we obtain
\be
\delta \vp(r) \sim \delta A r^{\frac{3}{2}} + \delta B+ \delta C\, r^4 e^{-\frac{33223}{31941}r}\,.
\ee
%where again we find a term $\delta B$ resulting from perturbing the constant term and an exponentially decaying piece multiplying the other missing parameter $\delta C$. Since neither term grows faster than the leading solution, both are kept and 
We see that for $\beta=-\nfrac{5}{9}$ the solution also contains three independent parameters.

The remaining pole of $p_3$ at $\beta=1$ is the result of the coefficients $h_1$ and $h_2$ diverging. These coefficients also diverge at $\beta=\pm\nfrac{1}{\sqrt{27}}$, but since $h_3$ does as well, this behaviour is not captured by $p_3$: we are able to multiply through by $(27\beta^2 - 1)$ in \eqref{diff eta} thereby removing this pole from the differential equation.
Nonetheless, the value $\beta=\pm\nfrac{1}{\sqrt{27}}$ is still problematic. In fact both $\beta=\pm\nfrac{1}{\sqrt{27}}$ and $\beta=1$ correspond to values at which the leading solution \eqref{solution} already breaks down.
The issue can be traced back to zeros occurring in the denominators on the right-hand side of the fixed point equation as discussed in sec. \ref{Exceptions}.

There are further values of $\beta$, corresponding to the zeros of the coefficients \eqref{coeffs f}, for which the differential equation \eqref{diff eta} is no longer correct. One example which can be seen straightforwardly is $\beta=0$. This has been already flagged up as a troublesome value in sec. \ref{Exceptions} and is actually another value for which the leading solution is already not valid.

\subsection{Numerical comparison}
\label{sec:numer-pow}
The authors of ref. \cite{Demmel2015b} tried matching their $\beta=1/6$ numerical solution to asymptotic behaviour given by:
\be 
\label{eq:fitansatz}
\vp(r) = A_2 \, r^2 \left(1 + u_1 \, r^{-1} + u_2 \, r^{-2}+\cdots \right) \, .
\ee
We have seen that this is not a valid asymptotic series for the fixed point equation \eqref{fp}, except at the special values $\beta_\pm$ as given in \eqref{beta n=2}, as in case (d). They also found no analytic match except at these special values, and  concluded that the asymptotic behaviour should be the result of a ``balanced regime'' which is taken to be $Ar^2$ but with logarithmic corrections. This bears some similarity to the asymptotic series we investigate in sec. \ref{sec:non-pow}, which however we will see in sec. \ref{sec:numer-non-pow} cannot provide the asymptotic solution because it does not have enough asymptotic parameters. Since the authors chose a value of $\beta$ that provides already $n_s=n_{ODE}=3$ constraints on the fixed point solution space through the fixed singularities, any number of asymptotic parameters less than the maximum $n_{ODE}=3$, will rule out a global solution. 

In this sense the authors struck lucky because we find a suitable power-law asymptotic solution with the maximum three parameters, namely \eqref{full-pow-beta1/6}.
The authors determined a fit of \eqref{eq:fitansatz} over the range $r\in[6,9]$. We can use this fit to
see how well our power-law asymptotic asymptotic solution \eqref{full-pow-beta1/6} does in matching their large $r$ behaviour as far as it was taken. As we will see, despite the very different leading behaviour at large $r$ we can find equally acceptable fits. 
\begin{figure}
\centering
\includegraphics[scale=0.4]{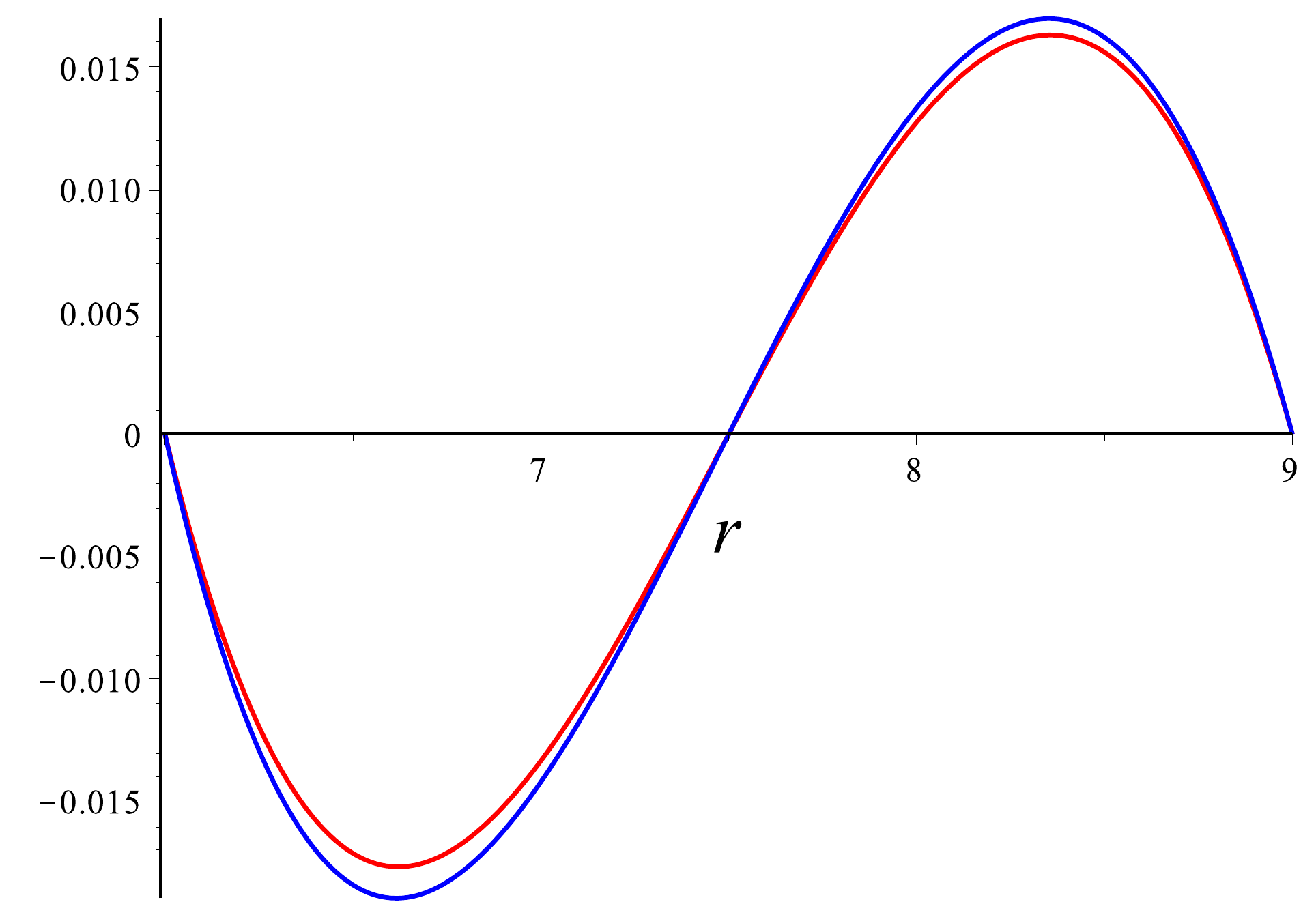}
\caption{Plot of the difference between \eqref{full-pow-beta1/6} and \eqref{eq:fitansatz}. The red curve uses \eqref{pow-param-sols1} and the blue curve uses \eqref{pow-param-sols2}.}
\label{fig:fit-pow}
\end{figure}

Their fit gave the solutions:
\be\label{eq:fitparamters}
\begin{split}
A^{\rm fit}_2 = & \, \phantom{-}\ 0.07705 \pm 0.00032 \, , \\
u_1^{\rm fit} = & \, -2.07514 \pm 0.05399 \, , \\
u_2^{\rm fit} = & \, -6.36855 \pm 0.25897 \, . 
\end{split}
\ee
Note that our asymptotic expansion \eqref{full-pow-beta1/6} to the level taken, is actually linear in $C$ and $\log(b)$, so it is straightforward to solve for these. Determining also $A$ by insisting that 
\eqref{full-pow-beta1/6} agrees with \eqref{eq:fitansatz} at $r=6$, 7.5 and 9, we find two solutions:
\begin{alignat}{2}
\label{pow-param-sols1}
A &= -5.6498\cdot10^{-5}\,, &\quad \log(b) =-4932.4\,, &\quad C =0.13864\,;\\
\label{pow-param-sols2}
A &=5.0025\cdot10^{-4}\,, &\quad\log(b) =3.6538\,,\ \ &\quad C=0.12571\,;
\end{alignat}
where the second seems more believable. On the other hand we note that the asymptotic expansion \eqref{full-pow-beta1/6}  suggests, but does not require,\footnote{for example we could rewrite the expansion in terms of $\log(b)$} that we apply it only to the region $r>b$, which would favour the first solution.
\begin{figure}
\centering
\includegraphics[scale=0.4]{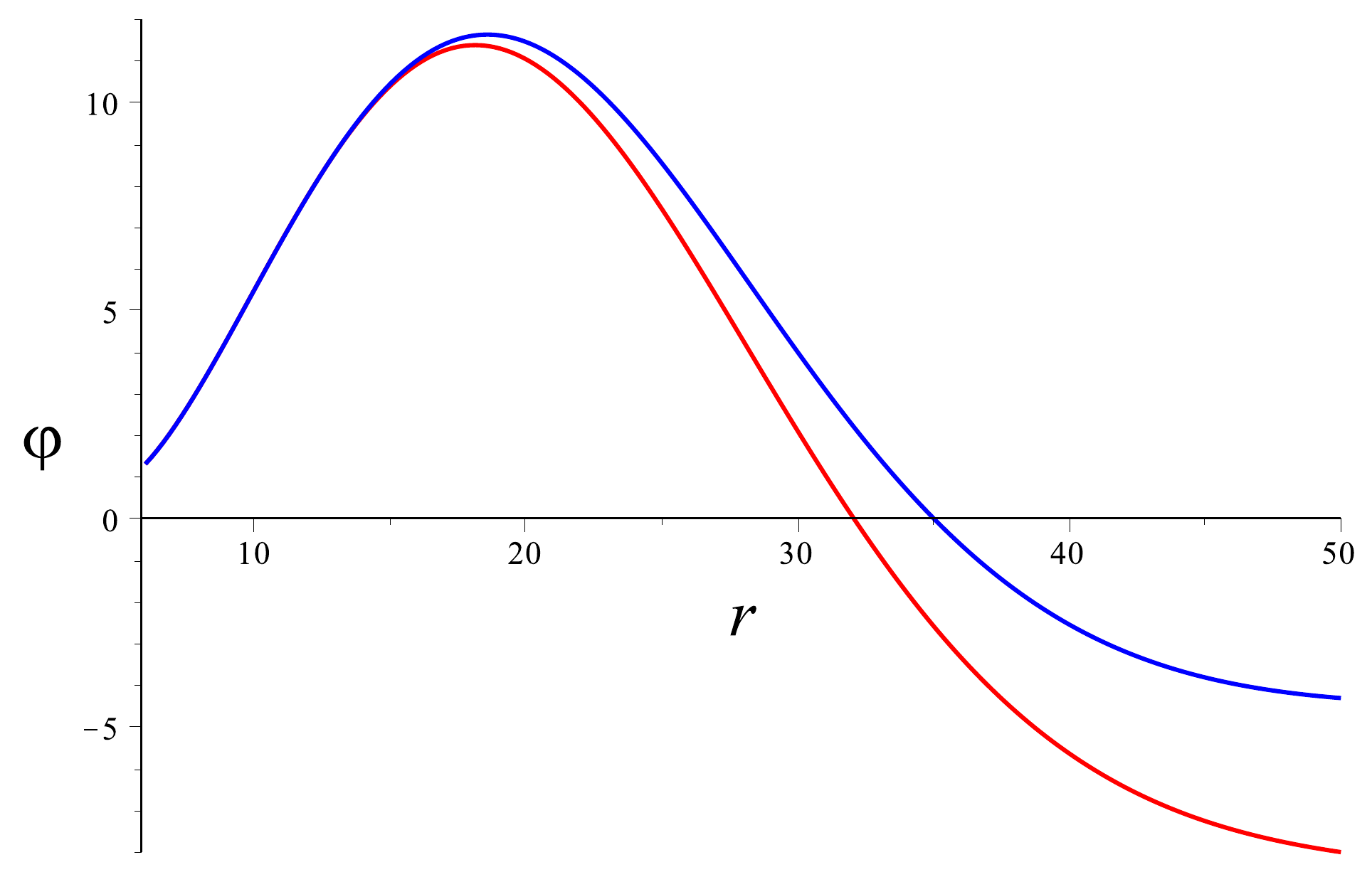}
\caption{The predicted large $r$ behaviour from the two fits. The red curve uses \eqref{pow-param-sols1} and the blue curve uses \eqref{pow-param-sols2}.}
\label{fig:fit-pow-asymptotic}
\end{figure}
It is not possible to distinguish by eye the solutions and the (fitted) numerical solution over the range $r\in[6,9]$, so instead we plot their difference in fig. \ref{fig:fit-pow} for the two possibilities \eqref{pow-param-sols1} and \eqref{pow-param-sols2}. As can be seen the error is almost the same in both cases and competitive with the error implied by the spreads in \eqref{eq:fitparamters}. Clearly the two possibilities   \eqref{pow-param-sols1} and \eqref{pow-param-sols2} are not both correct. Determining which, if either, is correct, would require computing the numerical solution to larger $r$. In particular note that the asymptotic solution \eqref{full-pow-beta1/6} fits  because the final term dominates in the fitted region, where it provides the $r^2$-like behaviour necessary to fit the data. Its exponential decay only becomes significant once $r>\sqrt{351}=18.73$ after which our fitted solutions peak and then turn negative, with the leading asymptotic parts of \eqref{full-pow-beta1/6} finally taking over, as can be seen in fig. \ref{fig:fit-pow-asymptotic}.

\section{Asymptotic expansion of a non-power law solution}
\label{sec:non-pow}

Power counting for $\vp(r)\sim r^n$ suggests that $\vp(r)\sim r^2$ should be the leading solution, since then the two sides of the fixed point equation, \eqref{lhs} and \eqref{rhs}, balance. 
However, as discussed just above sec. \ref{Exceptions},
this fails to be the case in general because it also happens that the left-hand side vanishes identically. Then the leading term on the right-hand side must also vanish, which is only true for specific values of $\beta$. The way out is analogous to that discussed in sec. \ref{exceptions}: since $r^2$ is annihilated by the left-hand side, we know that $r^2\log(r)$ will survive and furthermore give us a pure power $r^2$, which is what we will need in order to balance the $r^2$ power coming from taking ratios on the right-hand side. 

In fact for good measure we also tried the general ansatz $\vp(r)\sim r^n(\log r)^p$. Then one finds that on the right-hand side of \eqref{fp}, the first ratio $\sim r^2$ while the second ratio $\sim r^0$. Therefore balance is achieved only for $n=2$ and $p=1$, since then as we have just seen the left-hand side also behaves as $\sim r^2$.

%However, if we were to multiply the $r^2$ by $\log(r)$, it is still close enough to $r^2$ (since for the asymptotic limit $\log(r)$ grows slower than any power of $r$) but its derivatives bring down $r^{-1}$, which gives a non zero left-hand side in the fixed point equation.

\subsection{Leading behaviour}
\label{sec:non-pow-leading}
We now study  the leading behaviour for the $r^2\log r$ ansatz. In full the above argument implies that this leading term has to be of the form
\be\vp(r)\sim k_1 r^2 \log\left(\tfrac{r}{b}\right)\,.\label{ansatzlog}\ee 
%(or more generally, a function of $\log\left(\tfrac{r}{b}\right)$) yield a now non-vanishing $r^2$ leading term both in the right and the left-hand side of \eqref{fp}.
% Here, $k_1$ is a $\b$-dependent constant whose value we will fix later, and $\b$ is an arbitrary constant.\par
%It only remains to find the value of $k_1$. For that, we expand the fixed point equation asymptotically in $r$ so the $k_1$ coefficient is then fixed by demanding that the leading term ($\sim r^2$) cancels. It reads,
Demanding that these agree, we find that $k_1$ is determined:
\be\label{non-pow k1} k_1=\dfrac{1-72\b+156\b^2}{9216\pi^2}\,,\ee
while $b$ is left undetermined.

\subsection{Sub-leading behaviour}
\label{sec:non-pow-subleading}
For the next  terms, it is easier to first write
$\vp(r)=r^2f(r)$ for some function $f$ and then change  variables $\nfrac{r}{b}\mapsto e^x$ such that $\vp(r)\mapsto e^{2x}f(x)$. Finally we divide by $r^2\equiv e^{2x}$ to simplify the fixed point equation \eqref{fp}. Once we do this, we can expand the fixed point equation in small exponentials in the following way
\be 0=f_a+f_b e^{-x}+f_c e^{-2x}+\cdots \,.\label{expapp}\ee
Here $f_b,f_c\dots$ give corrections that are smaller than any power so for the moment we can discard these pieces and concentrate on 
% anticipating that we will find a powers of $1/x$ in the asymptotic expansion, we can keep working only with with 
$f_a$, which has the following expression
\beal f_a&=2 f'(x)-\frac{30 \beta  (2 \beta -1) (6 \beta -5) \left(2 f''(x)+5 f'(x)+2 f(x)\right)}{4608\pi ^2 \left((3 \beta -1) f'(x)+(6 \beta -5) f(x)\right)}-\nonumber\\
&-\frac{\beta  (1-6 \beta )^2 \left(2 (9 \beta +5) f'''(x)+(63 \beta +31) f''(x)+(63 \beta +25) f'(x)+6 (3 \beta +1) f(x)\right)}{4608\pi ^2 \left((3 \beta +1)^2 f''(x)+(3 \beta  (9 \beta +5)+1) f'(x)+6 \beta  (3 \beta +1) f(x)\right)}\,.\label{EqaA}\eeal
In order to find the subleading terms, the procedure is as follows. Using \eqref{ansatzlog},  we find $f(x)=k_1 x$ so we can plug it in \eqref{EqaA}. The leading piece in  \eqref{EqaA} is thus given by a constant at large $x$, as already true of the first term in \eqref{EqaA}. We recover the coefficient $k_1$ by demanding that the constant part cancels.
%Since the leading piece in \eqref{EqaA} is given by the first derivative of $f$ our ansatz is such as its first derivative behaves as this leading term. Then we can substitute this in the equation and find the value of the coefficient by demanding that this terms cancels.\par
When doing so, we find that \eqref{EqaA} behaves now as $f_a\sim x^{-1}$ so we now add a term $k_2 \log\left({x}/{c}\right)$, where $c$ is an arbitrary constant, as indicated by the first term again. Demanding now the vanishing of the $x^{-1}$ coefficient in \eqref{EqaA} we find the value of $k_2$ (this, and the rest of the coefficients are listed in appendix \ref{kivalues}).
To find the next term we substitute 
\be f(x)=k_1 x+k_2 \log\left(\frac{x}{c}\right)\,,\ee
into \eqref{EqaA}, with the already known coefficients $k_1$ and $k_2$,
and thus find that the leading piece behaves now as ${\log(x)}/{x^2}$. This implies we add to our ansatz $k_3{\log\left({x}/{c}\right)}/{x}$ and find the value of $k_3$ that cancels this term. In the end, we find  $f(x)$ is given by
\beal f(x)&=k_1 x+k_2\log\left(\dfrac{x}{c}\right)+k_3\dfrac{\log\left(\dfrac{x}{c}\right)}{x}+\dfrac{k_4}{x}+k_5\dfrac{\log^2\left(\dfrac{x}{c}\right)}{x^2}+\cdots \,, \label{f}
\eeal
where the $k_i$ coefficients are given in the Appendix \ref{kivalues}.\par

It is worth noticing that we have a constant, $b$, in \eqref{ansatzlog} and another one, $c$, in \eqref{f}. The constant $b$ is also captured in $f$ by using the translation invariance of $f_a$. Thus if $f(x)$ is a solution, so is $f(x+x_0)$, where $x_0\equiv-\log(b)$. One might think that we already have two free parameters in the solution. However, this is not the case: it is easy to see, with the values of appendix \ref{kivalues}, that
\be \dfrac{\partial f(x+x_0)}{\partial x_0}=-\dfrac{k_1 c}{k_2}\dfrac{\partial f(x+x_0)}{\partial c}\, ,\ee
which implies that the two constants can be combined into one, and therefore there is actually only one independent parameter. Since we already know we can dispense with $b$ and then recover it by $x$-translation invariance, in the following we set $b=c=1$ when working with $f(x)$ and instead fold them into a parameter $A$ in the end for $\vp(r)$. Indeed, changing variables back to $r$, we see that the whole solution is then given in \eqref{vptotal}.
%\be \vp(r)=r^2\Bigg\{k_1\log\left(\dfrac{r}{A}\right)+k_2\log\left(\log\left(\dfrac{r}{A}\right)\right)+k_3\dfrac{\log\left(\log\left(\dfrac{r}{A}\right)\right)}{\log\left(\dfrac{r}{A}\right)}+\dfrac{k_4}{\log\left(\dfrac{r}{A}\right)}+k_5\dfrac{\log^2\left(\log\left(\dfrac{r}{A}\right)\right)}{\log\left(\dfrac{r}{A}\right)}+\cdots\Bigg\}\,.\label{vptotal}\ee

\subsection{Exceptions}\label{exceptionss}

By looking at the coefficients $k_i$ in Appendix \ref{kivalues} it can be seen that there are some values of $\b$ for which \eqref{vptotal} is not an acceptable solution. These are listed below, depending on their nature.

\paragraph{$\boldsymbol{\b=0}$ and $\boldsymbol{\b=\b_\pm}$:}
 for these values $\vp(r)=r^2\log\left({r}/{b}\right)$ is not a solution of the fixed point equation. This second case was expected, since we have seen in \eqref{beta n=2}  that for these values $\vp(r)\sim r^2$ is a solution, without the need to add the $\log$ term. Actually, what happens  in both cases is that  the leading power on the right-hand side decreases, so the asymptotic behaviour is dictated by the left-hand side, which vanishes for $\vp(r)\sim r^2$ but not for $\vp(r)\sim r^2\log r$. However for $\beta=0$ we saw that this exceptional behaviour left us then with only the $n=0$ solution \eqref{n=0no2}.

\paragraph{$\boldsymbol{\b=-\nfrac{1}{3}}$ and $\boldsymbol{\b=\nfrac{5}{6}}$:}
for these values the asymptotic solution is still of the form \eqref{vptotal}, or equivalently \eqref{f}, but because these coefficients cancel leading contributions in \eqref{fp} the coefficients $k_i$ have different values that do not correspond with just  substituting the above values of $\b$ into \eqref{kigeneral}. Following the same procedure, but with the right leading contributions for these cases, we find the new coefficients given in \eqref{ki13} for $\b=-\nfrac{1}{3}$, and \eqref{ki56} for $\b=\nfrac{5}{6}$. %which are listed in appendix \ref{kivalues}.

\subsection{Finding the missing parameters}
\label{missing params 2}

In order to find the two missing parameters, we can linearise \eqref{EqaA} about \eqref{vptotal}, writing  $f(x)\mapsto f(x)+\e\eta(x)$.   We get a differential equation for the perturbed function
\be
h_3x\eta'''(x)+h_2x\eta''(x)+ 2x^2\eta'(x)+h_0\eta(x)=0\label{eqa}\,,
\ee
where the $h_i$ are the following functions of $\b$ ($h_1=2$ so this simple value is substituted directly):
\beal
h_0&=\frac{4 \beta  \left(3 \beta  \left(9 \beta  \left(312 \beta ^2-334 \beta -5\right)+406\right)-29\right)+5}{3 \beta  (3 \beta +1) (6 \beta -5) (12 \beta  (13 \beta -6)+1)}\,,\label{h0nonpow}\\
%h_1&=2\,,\\
h_2&=\frac{-4104 \beta ^4+108 \beta ^3+642 \beta ^2-37 \beta +1}{3 \beta  (3 \beta +1) \left(156 \beta ^2-72 \beta +1\right) }\,,\label{h2nonpow}\\
h_3&=-\frac{2 (1-6 \beta )^2 (9 \beta +5)}{3 (3 \beta +1) \left(156 \beta ^2-72 \beta +1\right) }\,.\label{h3nonpow}
\eeal
We know one solution to this equation is 
\be 
\label{f-prime}
\eta=\frac{\partial f}{\partial x} = k_1 + \frac{k_2}x -k_3 \frac{\log(x)}{x^2}+\cdots\,,
\ee
by translation invariance.  The other two solutions cannot go like a power for large $x$ since this would make the $\eta'''$ and $\eta''$ terms subleading already. In other words for power-law solutions, \eqref{eqa} behaves like a linear first order differential equation, with thus (up to a scale) only one solution. We need to ansatz a solution that can make $\eta''$ and/or $\eta'''$ as important as the $\eta$ or $\eta'$ terms. This motivates trying 
\be 
\label{Lp-ansatz}
\eta= e^{L x^p}\,,
\ee 
with $L\ne0$ and $p>1$ .
%some positive number.\footnote{Notice that, in principle, p can be either positive or negative but not zero, since a constant is not a solution of \eqref{EqaA}} 
In that case, we have
\beal
\eta&= e^{L x^p}\,,\\
\eta&'= L p x^{p-1} e^{L x^p}\,,\\
\eta&''= L^2 p^2 x^{2 p-2} e^{L x^p}+L (p-1) p x^{p-2} e^{L x^p}\sim L^2 p^2 x^{2 p-2} e^{L x^p}\,,\\
\eta&'''= L^3 p^3 x^{3 p-3} e^{L x^p}+L^2 (p-1) p^2 x^{2 p-3} e^{L x^p}+L^2 p^2 (2 p-2) x^{2 p-3} e^{L x^p}+\nonumber\\
&\quad+L (p-2) (p-1) p x^{p-3} e^{L x^p}\sim L^3 p^3 x^{3 p-3} e^{L x^p}\,,	\\
\eeal
where we are keeping only the (asymptotically) leading terms.
Therefore, we end with
\be e^{L x^p}\left(h_0+2 L p x^{p+1}+h_2 L^2 p^2 x^{2 p-1}+h_3 L^3 p^3 x^{3 p-2}\right)=0\,.
\label{leadeq}\ee
Actually we can discard the $h_0$ term in this expression since for $p>1$, it can never be leading. Of the remaining terms, we want the leading terms to cancel against each other, so there are three options: 
\paragraph*{Option 1:} The second and third terms are the leading ones, then
\be
 p+1=2p-1\Rightarrow p=2\,,
 \ee
but for this value the last term will in fact be leading, so this is exlcuded.
\paragraph*{Option 2:} The last two terms are leading,
\be
2p-1=3p-2\Rightarrow p=1\,.
\ee
but for this value the second term will be leading, so this is exlcuded.
\paragraph*{Option 3:} The second and the last term are leading,
\be
p+1=3p-2\Rightarrow p=\dfrac{3}{2}\,.
\ee
This is allowed, since the third term is now subleading.
Now demanding that the leading term vanishes, implies that $L$ has to fulfill
$
2  +h_3 p^2 L^2=0\,,
$
%The trivial solution $L=0$ has to be discarded since 
\ie
\be L=\pm\dfrac{2 }{3}\sqrt{\dfrac{- 2}{h_3}}\label{Bsol}\,.\ee
This will take a real or imaginary value depending on the sign of $-h_3$.
%\begin{figure}
%{\centering
%\includegraphics[scale=1.2]{h1h3.pdf}
%\caption{$-\frac{2}{h_3}$ depending on $\b$.}}
%\end{figure}
Thus for $\b<-\nfrac{5}{9},\; -\nfrac{1}{3}<\b<\b_-$ and $\b>\b_+$, $L$ is real, where $\beta_\pm$ is defined in \eqref{beta n=2}. Then only the negative root in \eqref{Bsol} is acceptable (since the other one grows exponentially for $r\to\infty$) and we see that there are thus only two parameters in the asymptotic solution. 

Otherwise $L$ takes an imaginary value. In this case  \eqref{Lp-ansatz} has unit modulus, so to see whether it is allowed we need to go to next order to compare its behaviour with the leading $k_1 x$ term in  \eqref{vptotal}. Thus we now substitute
\be 
\label{Lp-ansatz-corrected}
\eta= e^{L x^{\frac{3}{2}} +\zeta(x)}\,,
\ee
where $\zeta$ grows slower than $x^{\nfrac32}$. In this way we find that 
\be 
\zeta= 4\frac{h_2}{h_3}x\,.
\ee
Now whether or not this perturbation is acceptable depends on the sign of $h_2/h_3$. We find that for $-\nfrac59<\beta<-\nfrac13$ the sign is negative. Therefore the perturbation is allowed. Together with the two parameters, the two solutions in \eqref{Bsol} get combined into real oscillatory combinations with an exponential tail provided by $\zeta$. We have therefore found a three parameter asymptotic solution.
For the other region of imaginary $L$, namely $\b_-<\b<\b_+$, we find that $h_2/h_3>0$. Thus in this region \eqref{Lp-ansatz-corrected} is an exponentially growing perturbation and is excluded. Therefore in this region of $\beta$, the asymptotic solution only has the one parameter $A$ in \eqref{vptotal}.

%There is also the possibility of $p<0$. However for this values the equation that we have to consider reads,
%
%\be e^{L x^p}\left(h_0+2 L p x^{p+1}+h_2 L (p-1) p x^{p-1}+h_3 L (p-2) (p-1) p x^{p-2}\right)=0\ee
%
%and the vanishing of the leading terms will require $p=0$, which contradicts the assumption of $p<0$ so there is no solution for negative p.\par

As in sec. \ref{Missing parameters}, we note that where these perturbations are allowed we can replace their linearised coefficients with full coefficients, since the perturbations can already be made as small as we like compared to the leading terms by increasing $x$.
We can summarise the full asymptotic solutions we have found so far as \eqref{non-pow-only}--\eqref{non-powOsc32}.

%, and recall that the $A$ parameter can be recovered by translating $x\mapsto x-\log A$.
%To summarise, the asymptotic solutions depending on the value of $\b$ are shown in the following table\par
%
%\begin{table}[htbp]
%\centering
%\begin{tabular}{ >{\centering\arraybackslash}m{0.45\textwidth} | >{\centering\arraybackslash}m{0.57\textwidth} }
%$\eta(x)$& $\b$ \\ 
%\hline
%\rule{0pt}{5ex} 
%$\d B e^{-\tfrac{2 }{3}\sqrt{\tfrac{-2 h_1}{3h_3}}x^{\frac{3}{2}}}$ & $\left(-\infty,-\tfrac{5}{9}\right)\cup\left( -\tfrac{1}{3},\tfrac{18-\sqrt{285}}{78}\right) \cup\left(\tfrac{18+\sqrt{285}}{78},\infty\right)\backslash\left\{0,\tfrac{5}{6}\right\}$\\ \vspace{1mm}
%\shortstack{$\d B e^{\tfrac{2 i }{3}\sqrt{\tfrac{2 h_1}{3h_3}}x^{\frac{3}{2}}}+\d C e^{-\tfrac{2 i }{3}\sqrt{\tfrac{2 h_1}{3h_3}}x^{\frac{3}{2}}}\equiv$\\
% $\d B \cos\left(\tfrac{2  }{3}\sqrt{\tfrac{2 h_1}{3h_3}}x^{\frac{3}{2}}\right)+\d C \sin\left(\tfrac{2  }{3}\sqrt{\tfrac{2 h_1}{3h_3}}x^{\frac{3}{2}}\right)$}&  $\left(-\tfrac{5}{9},-\tfrac{1}{3}\right)\cup\left(\tfrac{18-\sqrt{285}}{78},\tfrac{18+\sqrt{285}}{78}\right)\backslash\left\{\tfrac{1}{6}\right\} $\\
%\noalign{\smallskip}
%\end{tabular} 
%\caption{Missing parameters as function of $\b$}
%\label{table1}
%\end{table}

\subsection{Exceptions}
\label{sec:missing2exceptions}
%\addtocontents{toc}{\setcounter{tocdepth}{2}} 
There are several values of $\b$ for which the differential equation \eqref{eqa} is not valid. The first two below relate to exceptions already considered in sec. \ref{exceptionss}, where we saw that the expansion coefficients in \eqref{vptotal} get altered. The remaining cases are caused  by the vanishing of one of the $h_i$ coefficients, as listed in \eqref{h0nonpow}--\eqref{h3nonpow}. This means that the corresponding term in \eqref{eqa} gets replaced by a term which grows more slowly at large $x$. Since the $h_0$ term played no r\^ole in the above analysis, exceptions arise only from the vanishing of $h_3$ and $h_2$.

% We can distinguish between those that were already posted in sec. \ref{exceptionss}, for which the ansatz is different in the first place, and those for which some of the $h_i$ coefficients vanish. 

%In the latter case what happens is that the leading order for the third derivative coefficient is lower than in the previous general case. In particular, for the values that vanish the $h_3$ coefficient, $\b=-\dfrac{5}{9}$ and $\b=\dfrac{1}{6}$, we need to go to higher order of the exponential expansion \eqref{expapp} to find the first non-vanishing coefficient for the third derivative term. 

\subsubsection{Altered coefficients}
\label{sec:altered}

\paragraph{$\boldsymbol{ \beta=-1/3}$:}
for this value, using the coefficients in \eqref{ki13} we can follow the same procedure to end up with the equation
\be\label{eqa3}
-\frac{9x^2 }{17}\eta'''(x)-\frac{45 x^2 }{34}\eta''(x)+2x^2\eta'(x)+\frac{775 }{238 }\eta(x)=0\,.\ee
Again, we know one solution is \eqref{f-prime}, using however the coefficients $k_i$ from \eqref{ki13}. Then the leading behaviour of the other two solutions does not involve the undifferentiated $\eta$. Indeed, dividing by $x^2$, the other three terms on their own give a differential equation with constant coefficients which is therefore solved with $\eta = e^{Lx}$, while the undifferentiated $\eta$ term then contributes $\sim e^{Lx}/x^2$, which can be neglected at leading order. $L$ thus solves a cubic. 
Discounting the $L=0$ solution (which is \eqref{f-prime} in disguise), we are left with a quadratic whose roots are $L=-L_\pm$, where
\be 
\label{Lpm}
L_\pm = \frac54\pm\frac{\sqrt {769}}{12}\,.
\ee
Since $-L_->0$ and $-L_+<0$, we discard the $-L_-$ solution and are  left with the two parameter asymptotic solution \eqref{non-powExpL}.
%\be 
%f_{asy}(x)+B\, e^{L_-x}\,,
%\ee
%(where recall $x\mapsto x-\log A$).

%We try again the ansatz $e^{Lx^p}$ (with $p>0$) so, at leading order, we end up with
%
%\be
%e^{L x^p}\left(h_0+2 L p x^{p+1}+h_2 L^2 p^2 x^{2 p}+h_3 L^3 p^3 x^{3 p}\right)=0\,.
%\ee
%Matching coefficients, the only acceptable solution is $p=\dfrac{1}{2}$. Moreover, now L is always real, so we only have a one-parameter solution corresponding to the $-$ sign.
%
%\be \d Be^{-\tfrac{4}{3} \sqrt{17}x^{\frac{1}{2}}}\,.\ee

\paragraph{$\boldsymbol{\b={5}/{6}}$:}
as before, but using now \eqref{ki56} the equation reads
\be-\frac{50x }{21 }\eta'''(x)-\frac{688x }{105 }\eta''(x)+2x^2 \eta'(x)+ \frac{388 }{105 }\eta(x)=0\,.\ee
This has the same form that \eqref{eqa} so trying the same ansatz it has the same solution for $p=\nfrac{3}{2}$ and thus we find the two-parameter asymptotic solution \eqref{non-powExp32alt}.
%\be \displaystyle 
%f_{asy}(x) +B \exp\left\{-\tfrac{2\sqrt{21}}{15}\,x^{\frac{3}{2}}\right\}\,,\qquad x\mapsto x-\log A\,.\ee
\paragraph{}The remaining possible exceptional values can be arranged according to which coefficient of \eqref{eqa} they cause to vanish.

\subsubsection{Third derivative}
\label{sec:third derivative}

There are two values that make $h_3$ vanish. For both of them, it is not that we have to go to the next order in \eqref{EqaA} to get the leading term, but that the third derivative term vanishes identically there. We need to go to higher order in the exponential expansion.

\paragraph{$\boldsymbol{\b=1/6}$:} In this case, in the exponential expansion of the fixed point equation,
\be f_a+f_b e^{-x}+f_c e^{-2x}+\cdots\,,\ee
not only does the third derivative term vanish in $f_a$ but also in $f_b$. In order to find the coefficient $h_3$ we need to consider $f_c$. The resulting equation is
\be \frac{78 x e^{-2 x}}{5 }\eta'''(x)-2x\eta''(x)+2x^2\eta'(x)+\frac{19 }{4}\eta(x)=0\,.\ee
An ansatz of the form $e^{Le^{2x}}$ provides the perturbation that involves the third derivative, by balancing against the second derivative part, with the rest subleading. But we see that $L=5/78$, which being positive, rules this out of the asymptotic series. The other perturbation is found by neglecting the third derivative term. In this case we get,
\be-2x \eta''(x)+2 x^2 \eta'(x)+\frac{19 }{4}\eta(x)=0\,.\ee
With an ansatz $e^{L x^p}$, 
%this yields
%\be e^{Lx^p}\left(\frac{19}{4}+2pL x^{p+1}-2p(p-1)Lx^{p-1}-2p^2L^2x^{2p-1}\right)=0\,\ee
%whose 
one finds the asymptotic solution $p=2,\, L={1}/{2}$. Again, this growing perturbation is ruled out in the asymptotic series, so we end up with only the one parameter solution $f_{asy}(x-\log A)$, \ie $\vp(r)$ as in \eqref{vptotal}.

\paragraph{$\boldsymbol{ \beta=-5/9}$:}
now the coefficient for $\eta'''$ appears in $f_b$ and we get
\be-\frac{4563xe^{-x}}{2407 }  \eta'''(x)-\frac{23056x}{12035 }\eta''(x)+2x2 \eta'(x)+\dfrac{103627 }{24070}\eta(x)=0\,.\ee
Thus similar to the previous case, an ansatz of the form $e^{Le^{x}}$ provides the perturbation that involves the third derivative. Since then $L=-\nfrac{23056}{22815}<0$ this rapidly decaying perturbation provides one of the missing parameters. Neglecting the third derivative term we get a similar equation to the previous case, for which the missing perturbation is $e^{L x^p}$, with again  $p=2$ but now $L=\nfrac{12035}{23056 }$. This is therefore still an exponentially growing perturbation and thus ruled out. Therefore in this case we have the two parameter asymptotic solution \eqref{non-powExpExp}.
%\be 
%f_{asy}(x-\log A) + B\, e^{-\frac{23056}{22815}\,e^x}\,.
%\ee
%(Note that translation invariance does not extend to the exponential corrections \eqref{expapp}.)

\subsubsection{Second derivative}\label{k2sec}

The coefficient $h_2$ vanishes for the two real roots of the quartic in \eqref{h2nonpow}, \cf table \ref{table2}.
% \ie $\b =$ $-0.4111$ and $0.3800$.
The differential equation reads now (with a new $\eta''$ term and coefficient $h_2$):
\be h_3 x \eta'''(x)+h_2 \eta''(x)+ 2 x^2 \eta'(x)+ h_0 \eta(x)=0\,.\ee
where the coefficients are also in the table. 
\begin{table}[htbp]
\centering
%	{\fontsize{13pt}{13pt}\selectfont
\begin{tabular}{c|c|c|c|c|c}
 $\b$ & $h_0$ %& $h_1$ 
 & $h_2$ & $h_3$& $L$ & $q=-\frac34-\frac{h_2}{2h_3}$ \\ 
\hline
%\rule{0pt}{5ex} 
$-0.4111$ &$4.1866 $%&$2$
&$2.1950$ & $0.7840$ & $\pm  1.0648 i$  & -2.1499\\
$0.3800$ &$5.8825$%&$2$
& $-7.6628$& $ 1.1209$ & $\pm 0.8905 i$ & 2.6681\\
\noalign{\smallskip}
\end{tabular} %}
\caption{Parameters for the differential equation and solutions, in the case that \eqref{h2nonpow} vanishes.}
\label{table2}
\end{table}
Comparing to the general case, \eqref{eqa}, we see that the only structural difference is that the $\eta''$ is now even more subleading. Since it actually played no r\^ole in the general case in determining the (formally) leading behaviour, the same ansatz \eqref{Lp-ansatz} solves this case and thus we find $L$ is given by \eqref{Bsol} but with $h_3$ as given in table \ref{table2}, and thus $L$ takes the imaginary values also listed in that table. Therefore as we saw in the general case, to determine whether this perturbation survives we need to go to the next order. Substituting \eqref{Lp-ansatz-corrected} we find that this time it is solved to leading order by
$e^\zeta = x^q$, where $q$ is also given in the table. Since overall the perturbation must grow slower than $k_1x$, the leading term in \eqref{vptotal}, we see that the two perturbations are excluded for $\beta=0.3800$ and thus we have only the one parameter solution \eqref{non-pow-only},
%$f_{asy}(x-\log A)$, 
while for $\beta=-0.4111$ we have the three-parameter solution
\eqref{non-powOsc}.

\subsection{Numerical comparison}
\label{sec:numer-non-pow}

From sec. \ref{sec:dimensionality-ef}, we already know that the relevant solution for $\beta=1/6$, namely the unadorned \eqref{vptotal}, cannot be the asymptotic limit of the numerical solution found in ref. \cite{Demmel2015b}, since we saw that its one free parameter is overconstrained. We can also see directly that the numerical solution, equivalently \eqref{eq:fitansatz} with \eqref{eq:fitparamters}, cannot match.
Using \eqref{kigeneral} we find at $\beta=1/6$:
\be 
k_1=-{\frac {5}{6912\,{\pi }^{2}}}, \
k_2=-{\frac {95}{55296\,{\pi }^{2}}}, \
k_3=-{\frac {1805}{442368\,{\pi }^{2}}}, \
k_4=-{\frac {95}{442368\,{\pi }^{2}}}, \
k_5={\frac {34295}{7077888\,{\pi }^{2}}}
\ee
Since the expansion only makes sense for $r\gg A$, we see the asymptotic solution implies that at large $r$, we have $\vp<0$ with $|\vp|$ growing faster than $r^2$, which is qualitatively different from the numerical solution.

\section{Discussion and Conclusions}
\label{sec:conclusions}

For the reasons set out in the Introduction, it is important to study functional truncations of the Wilsonian effective action, where full functions worth of couplings are kept.
Despite the complicated nature of the corresponding fixed point equations, in particular for the $f(R)$ approximation which then leads to a non-linear second or third order ODE for the corresponding scaled quantity $\vp(r)$, we have seen that by adopting techniques first developed in \cite{Morris:1994ki,Morris:1994ie,Morris:1994jc} and applied to this area in  \cite{Dietz:2012ic,Dietz:2016gzg}, it is reasonably straightforward to extract general key properties of the solutions, through an asymptotic analysis. The corresponding asymptotic solutions are set out as a summarised list in sec. \ref{sec:asymptotics-overview}, where also links are provided to the subsections where these are derived.

In particular, before resorting to a laborious numerical treatment, one can map out the dimensionality of the fixed point solution spaces using the counting formula \eqref{counting}. These spaces divide into sets depending on the number of free parameters, $n_{asy}$, in the corresponding  asymptotic  solution. We saw examples of this in sec. \ref{sec:dimensionality-ef}.
Finding the asymptotic solutions together with their complete set of free parameters, is thus key to this, as it is in fact for validating any numerical solution (as discussed in sec. \ref{sec:validate}) since without matching to an asymptotic solution one can never be sure that the hoped-for global numerical solution does not end in a moveable singularity at some large $r$. Moreover a full knowledge of the asymptotic behaviour provides insight and guidance for developing the numerical solution.
We provide an example of this in sec. \ref{sec:numer-pow} where we match the relevant asymptotic solution to the numerical solution found in ref. \cite{Demmel2015b}, see also sec. \ref{sec:numer-non-pow}.

In the original applications \cite{Morris:1994ki,Morris:1994ie,Morris:1994jc,Morris:1995he,Morris:1997xj}, one immediately found the (unique) leading behaviour of the asymptotic solution since this was simply given by scaling dimensions, \viz \eqref{Vscale}, neglecting the complicated part of the fixed point equation that describes the quantum corrections. In functional truncations for quantum gravity, it is now clear that this is typically no longer the case, as discussed in sec. \ref{sec:quantum}. Instead the quantum corrections remain important no matter how large the curvature $R$ is taken, for readily identifiable physical reasons.\footnote{The same was found to be true for metric in the conformal truncation of ref. \cite{Dietz:2016gzg}.} 

Thus a little sleuthing is required to find all possible leading terms for an asymptotic solution in functional truncations to quantum gravity. The strategy, as set out in sec. \ref{Leading behaviour}, is to start with a general ansatz, figure out which terms in the fixed point equation are then the most important at large $r$ and then require that these terms balance, \ie that these leading pieces cancel amongst themselves. The  possible ans\"atze are actually quite limited because most of any function $\vp(r)$ can be neglected in the large $r$ limit. In the example fixed point equation we chose, namely the ODE \eqref{fp} from ref. \cite{Demmel2015b}, we tried power law $\vp(r)\sim r^n$ as explained in sec. \ref{Leading behaviour}, resulting in solutions $n=0,$ $3/2$ and $n_\pm(\b)$ as summarised in cases (a) to (e) in sec. \ref{sec:asymptotics-overview}.
%\footnote{Ref. \cite{Dietz:2016gzg} provides another more involved example of power-law asymptotics for a fixed point equation where the right-hand side is only expressed implicitly through a momentum integral.} 
We also tried $\vp(r)\sim r^n (\log r)^p$, finding just the one solution, $n=2$ with $p=1$, that is presented in sec. \ref{sec:non-pow} and summarised as case (f) in sec. \ref{sec:asymptotics-overview}. Already this more complicated leading asymptotic solves the equation only through special circumstances, as explained at the beginning of  sec. \ref{sec:non-pow}. 

Carefully considering exceptions that appear in various regions, and at various special points of the endomorphism parameter $\beta$, including in sub-leading terms that we are about to discuss, we furnish a total of 15 different asymptotic series in sec. \ref{sec:asymptotics-overview}. In fact as shown in sec. \ref{exceptions}, there are further modifications of \eqref{solution} at discrete values of $\beta$ signalled by divergences in one of the subleading coefficients, potentially countably infinite in number.

Developing the leading asymptotic into a series $\vp_{asy}(r)$, complete with sub-leading corrections, is  the most straightforward part of the procedure, \cf secs. \ref{Sub-leading behaviour} and \ref{sec:non-pow-subleading}. However if the asymptotic series has $n_{asy}<n_{ODE}$ free parameters ($n_{ODE}$ being the order of the ODE),  we cannot be sure we have found the full asymptotic series until we have understood where the `missing' parameters have gone. This is where we see another huge difference \cite{Dietz:2012ic,Dietz:2016gzg} from the early applications \cite{Morris:1994ki,Morris:1994ie,Morris:1994jc,Morris:1995he,Morris:1997xj}. There it was always the case that $n_{asy}=1$ while $n_{ODE}=2$. The missing parameter always corresponded to a perturbation that grew rapidly, faster than the leading term in the asymptotic series, and thus could not be added without invalidating it. This perturbation could be understood to be the linearised expression of moveable singularities in the ODE. On the contrary here it is typically the case that the full asymptotic series contains further free parameters. It is clear that this is another expression of the fact that the quantum corrections do not decouple in the large $r$ limit. 

Finding these parameters, or proving that they are legitimately excluded, can be straightforwardly achieved through the following strategy. We perturb the asymptotic solution, writing $\vp(r)=\vp_{asy}(r)+\zeta(r)$, and keep only terms linear in $\zeta$. The result is a linear ODE, which is typically simple, since in the coefficients we only need the leading terms at large $r$. The task is further simplified since we are only looking for the leading behaviour of the solutions $\zeta$, and since for every parameter $a$ in $\vp_{asy}(r)$ we already know that: 
\be 
\zeta(r) = \zeta_{a}(r) := \frac{\partial}{\partial a}\vp_{asy}(r)
\ee
is a solution.
To find the  solutions, $\zeta=\zeta_m(r)$ corresponding to the missing parameters, the easiest way is to find an ansatz which can balance different terms in the, now linear, ODE. With a little thought it is always possible to find all $n_{ODE}$ solutions. A helpful hint is provided by noting that the highest derivatives must  have a r\^ole to play in at least one of the solutions. Once we have found $n_{ODE}$ linearly independent solutions, we are ready to classify them. If they grow faster than the leading term in $\vp_{asy}(r)$, they have to be discarded, as explained above. On the other hand if they grow slower than this leading term, we can add them to the asymptotic series with a \emph{finite} coefficient. This is because we can always take $r$ large enough to make the linearisation step valid, whatever size of coefficient we take. In this paper we provide numerous examples of this procedure in secs. 
\ref{Missing parameters}, \ref{missing params 2} and \ref{sec:missing2exceptions}, culminating in 11 different full asymptotic series in cases (e) and (f) in sec. \ref{sec:asymptotics-overview}, and a zoo of different $\zeta_m$, including powers of $r$, exponentials of $-r$ or $-r^2$, and $\sin\log r$ type terms. Needless to say, finding these missing terms is also important for matching to numerical solutions  \cite{Dietz:2012ic,Dietz:2016gzg}. We saw in sec. \ref{sec:numer-pow} that matching to the
 numerical solution found in ref. \cite{Demmel2015b}, crucially relied on the $C\, r^2 e^{-\nfrac{r^2}{351}}$ term in \eqref{full-pow-beta1/6}. Matching at high accuracy these full asymptotic solutions to numerical solutions, requires developing the asymptotic series, complete with the new parameters, to higher order. We do not do this in this paper, but examples can be found in refs. \cite{Dietz:2012ic,Dietz:2016gzg}, where we see that the non-linear parts of the ODE then generate sub-leading terms involving all the parameters. 

Although the eigenoperator spectrum was not addressed in this paper, asymptotic techniques were developed for them also  \cite{Morris:1994ie,Morris:1994jc,Morris:1996xq,Morris:1996nx,Bridle:2016nsu} and applied to asymptotic safety in  \cite{Dietz:2012ic,Dietz:2016gzg,Benedetti:2013jk}.

In sec. \ref{sec:dimensionality-ef} we used the above full asymptotic series to map out the dimensions of the solution spaces for different values of the endomorphism parameter $\beta$. This endomorphism parameter, together with the other one $\alpha$ which was ultimately set to $\beta-2/3$, was introduced to provide extra flexibility in designing the way modes are integrated out in the flow equations \cite{Demmel2015b}, in particular with the aim of ensuring that for some value of this parameter there is an isolated fixed point solution suitable for building an asymptotically safe theory of quantum gravity. 
Much the same strategy has also been followed in refs. \cite{Demmel:2014hla,Ohta:2015efa,Ohta2016,Falls:2016msz}. Such a freedom would indeed appear to be inherent in exact RG descriptions of quantum gravity, so it is certainly important to explore its consequences. However as we have seen in sec. \ref{sec:dimensionality-ef}, the freedom to change this parameter opens a Pandora's box. Depending on the value of $\beta$ and the asymptotic behaviour, there are no solutions, discrete fixed points, lines, or planar regions of fixed points. We discussed these briefly in sec. \ref{sec:which} in the light of results elsewhere in the literature. As we saw in sec. \ref{sec:intro-phys}, $\vp_{asy}(r)$ provides the fixed point equation of state through the limit in eqn. \eqref{phys}. In sec. \ref{sec:phys}, we saw this led to several possible scenarios.

Since quantum fluctuations remain strongly coupled at large $r$, it is not surprising that the results are sensitive to the formulation.  However ultimately we would want to see universality expressed as qualitatively the same behaviour for the fixed point and the corresponding equation of state, independent of the details of the regularisation, providing the regularisation is not singular in some way. Clearly, further research is required to improve the approximations. Fortunately the asymptotic techniques explained in this paper, are sufficiently powerful to allow the solution of much more sophisticated approximations, for example cases where the right-hand side of the flow equation is awkward or impossible to evaluate exactly \cite{Dietz:2016gzg}.
Finally, applying the techniques we have described here to other formulations that have already been developed \cite{Percacci:2015wwa,Labus:2015ska,Eichhorn:2015bna,Demmel:2014fk,Demmel:2012ub,Demmel:2013myx,Benedetti:2013jk,Demmel:2014hla,Demmel2015b,Ohta:2015efa,Ohta2016,Percacci:2016arh,Falls:2016msz,Ohta:2017dsq}, will no doubt further elucidate the situation.

\section*{Acknowledgments}
It is a pleasure to thank Astrid Eichhorn and Jan Pawlowski for discussions.
TRM acknowledges support from STFC through Consolidated Grant ST/L000296/1. ZS acknowledges support through an STFC studentship. SGM acknowledges the support of the Spanish MINECO {\em Centro de Excelencia Severo Ochoa} Programme under grant SEV-2012-0249 and COST actions MP1405 (Quantum Structure of Spacetime) and COST MP1210 (The String Theory Universe).

\appendix

\section{Power law solution coefficients}
\label{AppendixA}

\begin{align}
k_2 =& \frac{1}{\beta ^2 (\beta  (3 \beta  (36 \beta  (15 \beta -22)+301)+2)-19)^2 (\beta  (3 \beta  (36 \beta  (21 \beta -37)+725)-164)-23)}\times\nonumber\\
&\times\Big(9 A \left(589824 \pi ^4 A^2 (\beta -1) \left(27 \beta ^2-1\right) (\beta  (3 \beta  (3 \beta  (9 \beta  (12 \beta  (9 \beta  (60 \beta -151)+1103)-3175)-\right.\nonumber\\
&\left.-4000)+4370)+908)-271)+\beta  (\beta  (3 \beta  (36 \beta  (15 \beta -22)+301)+2)-19)^2 \times\right.\nonumber\\
&\left.\times(\beta  (3 \beta  (54 \beta  (28 \beta -33)+473)+76)-35)\right)\,,\\
k_3 =&\frac{1}{\beta ^3 \left(324 \beta ^4-810 \beta ^3+636 \beta ^2-83 \beta -2\right) \left(1620 \beta ^4-2376 \beta ^3+903 \beta ^2+2 \beta -19\right)^3}\times\nonumber\\
&\times\dfrac{1}{ \left(2268 \beta ^4-3996 \beta ^3+2175 \beta ^2-164 \beta -23\right)}\Big(1152 \pi ^2 A^2 \left(294912 \pi ^4 A^2 \left(21664553744880 \beta ^{17}-\right.\right.\nonumber\\
&-131103093477600 \beta ^{16}+335182432132080 \beta ^{15}-465992520928740 \beta ^{14}+\nonumber\\
&+373012915696569 \beta ^{13}-160032473858853 \beta ^{12}+20341799162595 \beta ^{11}+\nonumber\\
&+10879448697531 \beta ^{10}-3992311992294 \beta ^9-184358900772 \beta ^8+235681642062 \beta ^7-\nonumber\\
&-3342432654 \beta ^6-9333036891 \beta ^5+381546579 \beta ^4+240424223 \beta ^3-15717769 \beta ^2-\nonumber\\
&-\left.2976296 \beta +275350\right)+\beta  \left(1620 \beta ^4-2376 \beta ^3+903 \beta ^2+2 \beta -19\right)^2\times\nonumber\\
&\times  \left(3670485840 \beta ^{11}-13593079800 \beta ^{10}+19462865328 \beta ^9-13229473554 \beta ^8+\nonumber\right.\\
&\left.\left.+3888160137 \beta ^7-11847951 \beta ^6-212450220 \beta ^5+13245732 \beta ^4+8143979 \beta ^3-752317 \beta ^2-\right.\right.\nonumber\\
&\left.\left.-148144 \beta +17570\right)\right)\Big)\, ,\\
k_4=&-\frac{1}{\beta ^4 \left(324 \beta ^4-2484 \beta ^3+2913 \beta ^2-500 \beta +7\right) \left(324 \beta ^4-810 \beta ^3+636 \beta ^2-83 \beta -2\right)}\times\nonumber\\
&\times\dfrac{1}{ \left(1620 \beta ^4-2376 \beta ^3+903 \beta ^2+2 \beta -19\right)^4 \left(2268 \beta ^4-3996 \beta ^3+2175 \beta ^2-164 \beta -23\right)}\times\nonumber\\
&\times\Big(2654208 \pi ^4 A^3 \left(262440 \beta ^7-691092 \beta ^6+579798 \beta ^5-139563 \beta ^4-21348 \beta ^3+5382+ \beta ^2\right.\nonumber\\
&\left.+1214 \beta -211\right) \left(294912 \pi ^4 A^2 \left(21664553744880 \beta ^{17}-131103093477600 \beta ^{16}+\right.\right.\nonumber\\
& +335182432132080 \beta ^{15}-465992520928740 \beta ^{14}+373012915696569 \beta ^{13}-\nonumber\\
&-160032473858853 \beta ^{12}+20341799162595 \beta ^{11}+10879448697531 \beta ^{10}-3992311992294 \beta ^9-\nonumber\\
&-184358900772 \beta ^8+235681642062 \beta ^7-3342432654 \beta ^6-9333036891 \beta ^5+381546579 \beta ^4+\nonumber\\
&+\left.240424223 \beta ^3-15717769 \beta ^2-2976296 \beta +275350\right)+\nonumber\\
&+\beta  \left(1620 \beta ^4-2376 \beta ^3+903 \beta ^2+2 \beta -19\right)^2 \left(3670485840 \beta ^{11}-13593079800 \beta ^{10}+\right.\nonumber\end{align}
\begin{align}
&+19462865328 \beta ^9-13229473554 \beta ^8+3888160137 \beta ^7-11847951 \beta ^6-212450220 \beta ^5+\nonumber\\
&+\left.\left.13245732 \beta ^4+8143979 \beta ^3-752317 \beta ^2-148144 \beta +17570\right)\right)\Big)\, .
\end{align}

Unfortunately the $k_5$ expression is too long to include in the paper. We also list the values of the $k_i$  for the special case $\beta=1/6$:
\beal 
k_1 &=-1296\,{\pi }^{2}{a}^{2}\,,\nonumber\\
k_2 &={\frac {27\,a \left( 3649536\,{\pi }^{4}{a}^
{2}+25 \right) }{20}}\,,\nonumber\\
k_3 &=-{\frac {1944\,{\pi }^{2}{a}^{2} \left( 
123254784\,{\pi }^{4}{a}^{2}+865 \right) }{25}}\,,\\
k_4 &=-{\frac {30233088\,{
\pi }^{4}{a}^{3} \left( 123254784\,{\pi }^{4}{a}^{2}+865 \right) }{125
}}\,,\nonumber \\
k_5 &=-{\frac {81\,a \left( 4034150189236224\,{\pi }^{8}{a}^{4}+
29839933440\,{\pi }^{4}{a}^{2}+18125 \right) }{4000}}\,.\nonumber
\eeal

\section{Non-power law solution coefficients}\label{kivalues}

Coefficients $k_i$ for general $\b$:
\beal k_1&=\frac{156 \beta ^2-72 \beta +1}{9216 \pi ^2}\nonumber\,,\\
k_2&=\frac{33696 \beta ^5-36072 \beta ^4-540 \beta ^3+4872 \beta ^2-116 \beta +5}{55296 \pi ^2 \beta  (3 \beta +1) (6 \beta -5)}\nonumber\,,\\
k_3&=\frac{\left(33696 \beta ^5-36072 \beta ^4-540 \beta ^3+4872 \beta ^2-116 \beta +5\right)^2}{331776 \pi ^2 (5-6 \beta )^2 \beta ^2 (3 \beta +1)^2 \left(156 \beta ^2-72 \beta +1\right)}\nonumber\,,\\
k_4&=\frac{2659392 \beta ^8-3044304 \beta ^7-449064 \beta ^6+971352 \beta ^5+8748 \beta ^4-67518 \beta ^3+4119 \beta ^2+235 \beta -25}{331776 \pi ^2 (5-6 \beta )^2 \beta ^2 (3 \beta +1)^2}\nonumber\,,\\
k_5&=-\frac{\left(33696 \beta ^5-36072 \beta ^4-540 \beta ^3+4872 \beta ^2-116 \beta +5\right)^3}{3981312 \pi ^2 \beta ^3 (3 \beta +1)^3 (6 \beta -5)^3 \left(156 \beta ^2-72 \beta +1\right)^2}\label{kigeneral}\,.
\eeal
Coefficients for $\b=-{1}/{3}$:
\be
\label{ki13}
k_1=\frac{17}{3456 \pi ^2}, \,
k_2=\frac{775}{96768 \pi ^2},\,
k_3=\frac{600625}{46061568 \pi ^2},\,
k_4=\frac{349525}{46061568 \pi ^2},\,
k_5=-\frac{465484375}{43850612736 \pi ^2},
\ee
and for $\b=5/6$:
\be 
k_1=\frac{1}{576 \pi ^2},\,
k_2=\frac{97}{30240 \pi ^2},\,
k_3=\frac{9409}{1587600 \pi ^2},\,
k_4=\frac{4171}{705600 \pi ^2},\,
k_5=-\frac{912673}{166698000 \pi ^2}.\label{ki56}
\ee

%\begin{flalign}
%&k_1=\frac{17}{3456 \pi ^2} \,,\nonumber&\\
%&k_2=\frac{775}{96768 \pi ^2}\nonumber\,,\\
%&k_3=\frac{600625}{46061568 \pi ^2}\nonumber\,,\\
%&k_4=\frac{349525}{46061568 \pi ^2}\nonumber\,,\\
%&k_5=-\frac{465484375}{43850612736 \pi ^2}\label{ki13}\,,\\[5pt]
%&\text{and for} \b=\dfrac{5}{6}:\nonumber\\[2pt]
%&k_1=\frac{1}{576 \pi ^2} \,,\nonumber&\\
%&k_2=\frac{97}{30240 \pi ^2}\nonumber\,,\\
%&k_3=\frac{9409}{1587600 \pi ^2}\nonumber\,,\\
%&k_4=\frac{4171}{705600 \pi ^2}\nonumber\,,\\
%&k_5=-\frac{912673}{166698000 \pi ^2}\label{ki56}\,.
%\end{flalign}

\bibliographystyle{hunsrt}
%\addbibresource{references.bib}
\bibliography{references}

\end{document}